\documentclass[11pt,draft]{IEEEtran2}
\usepackage{color}     
\usepackage{epsf,psfrag,amssymb,amsfonts,latexsym,cite,verbatim,enumerate,subfigure}
\usepackage{srcltx,amsmath,cases, graphicx}
\usepackage{times, multirow, array, soul}
\usepackage[mathscr]{eucal}
\usepackage[normalem]{ulem}  

\def\psfancypar#1#2{\begingroup\def\par{\endgraf\endgroup\lineskiplimit=0pt}
               \setbox2=\hbox{\large\sc #2}
               \newdimen\tmpht \tmpht \ht2 \advance\tmpht by \baselineskip
               \font\hhuge=Times-Bold at \tmpht
               \setbox1=\hbox{{\hhuge #1}}
               \count7=\tmpht \count8=\ht1
               \divide\count8 by 1000 \divide\count7 by \count8 
               \tmpht=.001\tmpht\multiply\tmpht by \count7 
               \font\hhuge=Times-Bold at \tmpht
               \setbox1=\hbox{{\hhuge #1}}
               \noindent
                \hangindent1.05\wd1
               \hangafter=-2 {\hskip-\hangindent
               \lower1\ht1\hbox{\raise1.0\ht2\copy1}%
                \kern-0\wd1}\copy2\lineskiplimit=-1000pt}

\newcommand{\Phibf}{\mbox{${\bf \Phi}$}}

\newcommand{\Psibf}{\mbox{${\bf \Psi}$}}

\newcommand{\E}{\mbox{{\rm E}}}

\def\boxit#1{\vbox{\hrule\hbox{\vrule\kern3pt
        \vbox{\kern3pt#1\kern3pt}\kern3pt\vrule}\hrule}}

\def\reals{ { {\rm  I \kern-0.15em R }  } }
\def\complex{ {\,{{\rm C} \kern-0.50em \raise0.20ex {  |}}\, }}

\def\mubf{\hbox{\boldmath$\mu$\unboldmath}}

\def\xibf{\hbox{\boldmath$\xi$\unboldmath}}

\def\Sigmabf{\hbox{$\bf \Sigma$}}

\def\Pibf{{\bf \Pi}}

\def\ebf{{\bf e}}
\def\fbf{{\bf f}}
\def\gbf{{\bf g}}
\def\hbf{{\bf h}}

\def\nbf{{\bf n}}

\def\pbf{{\bf p}}

\def\rbf{{\bf r}}
\def\sbf{{\bf s}}

\def\ubf{{\bf u}}
\def\vbf{{\bf v}}
\def\wbf{{\bf w}}
\def\xbf{{\bf x}}
\def\ybf{{\bf y}}

\def\rbf{{\bf r}}
\def\xbf{{\bf x}}
\def\ybf{{\bf y}}
\def\Abf{{\bf A}}
\def\Bbf{{\bf B}}
\def\Cbf{{\bf C}}
\def\Dbf{{\bf D}}
\def\Ebf{{\bf E}}
\def\Fbf{{\bf F}}
\def\Gbf{{\bf G}}
\def\Hbf{{\bf H}}
\def\Ibf{{\bf I}}

\def\Kbf{{\bf K}}

\def\Mbf{{\bf M}}

\def\Qbf{{\bf Q}}
\def\Rbf{{\bf R}}

\def\Tbf{{\bf T}}

\def\Wbf{{\bf W}}

\def\Cc{{\cal C}}

\def\Hc{{\cal H}}
\def\Ic{{\cal I}}

\def\Kc{{\cal K}}

\def\Nc{{\cal N}}

\def\Rc{{\cal R}}
\def\Sc{{\cal S}}

\def\be{\vskip .3cm \begin{equation}}
\def\ee{\end{equation} \vskip .4cm \noindent}
\newcommand{\R}{\mbox{$\hat {\bf R}_{N}$}}

\def\Rxx{\Rbf_{\ssstyle X\kern-.1em X}}

\let\ssstyle=\scriptscriptstyle

\def\Kout{\setbox1=\hbox{\Huge\bf K}\hbox to
1.05\wd1{\hspace{.05\wd1}
}}
\def\Sout{\setbox1=\hbox{\Huge\bf S}\hbox to 1.05\wd1{\hspace{.05\wd1}
}}

  \ifx\LabelFigloaded\MYundefined\relax
  \else
   \fi

  \def\LabelFigloaded{\relax}

  \chardef\LabelFigCatAt\the\catcode`\@
  \catcode`\@=11

 \let\LabelFigwlog@ld\wlog
 \def\wlog#1{\relax}

 \ifx\\\MYundefined@
    \let\\\relax
 \fi

  \def\ms@g{\immediate\write16}

 \def\N@wif{\csname newif\endcsname }
 \def\Temp@ {\N@wif\ifIN@}
 \ifx\INN@\MYundefined@
    \else \let\Temp@\relax
 \fi
 \Temp@

  \def\IN@{\expandafter\INN@\expandafter}
  \long\def\INN@0#1@#2@{\long\def\NI@##1#1##2##3\ENDNI@
    {\ifx\m@rker##2\IN@false\else\IN@true\fi}%
     \expandafter\NI@#2@@#1\m@rker\ENDNI@}
  \def\m@rker{\m@@rker}
 
  \newtoks\Initialtoks@  \newtoks\Terminaltoks@
  \def\SPLIT@{\expandafter\SPLITT@\expandafter}
  \def\SPLITT@0#1@#2@{\def\TTILPS@##1#1##2@{%
     \Initialtoks@{##1}\Terminaltoks@{##2}}\expandafter\TTILPS@#2@}

 \def\Shifted@@#1#2#3{\setbox0=\hbox{#3}%
   \raise -\dp0\vbox {\kern-#2%
       \hbox {\kern#1\unhbox0\kern-#1}%
           \kern#2}}

 \newcount\gridcount
 \newbox\auxGridbox@ \newbox\hGridbox@ \newbox\vGridbox@
 \newbox\Labelbox@ \newbox\auxLabelbox@
 \newbox\Coordinatebox@
 \newtoks\Labeltoks@
 \newdimen\Wdd@ \newdimen\Htt@
 \newdimen\Wddd@ \newdimen\Httt@
 
 \def\Wr@{\immediate\write16}

 \newdimen\GL@wd
 \def\GridLineWidth#1{\GL@wd=#1}

 \def\gobble#1{}
 \def\EdgeErr@{\Wr@{}%
      \Wr@{\string\Edges\space argument
      1, 10, 100 or 1000 please\string!}%
      }

 \newcount\Edgect@

 \def\Sweepup#1\endSweepup{}

 \def\SetEdges@{%
    \edef\Zr@@s{\expandafter\gobble\number\Edgect@\empty}%
        \count255=0\Zr@@s\relax
        \ifnum\count255=\z@\else\EdgeErr@\show\tailtest\fi
        \count255=1\Zr@@s\relax
        \ifnum\count255=\Edgect@\relax\else\EdgeErr@\show\leadtest\fi
    \EdgGl@b\edef\Zr@s{\expandafter\gobble\Zr@@s\empty}
    \ifnum\Edgect@>\@ne\relax\EdgGl@b\let\L@Dc\empty
        \else\EdgGl@b\edef\L@Dc{\string.}\fi
    \ifnum\Edgect@>\@ne\relax
        \EdgGl@b\edef\Edgescale@##1{\divide##1 by \Edgect@}%
        \else\EdgGl@b\edef\Edgescale@##1{}\fi
    }

 \def\Edges#1{\Edgect@=#1\relax
     \let\EdgGl@b\global \SetEdges@}

 \Edges{1}

 \def\hhrule{\hrule height \GL@wd\vskip-.\GL@wd}

 \def\hRule@{%
   \advance\gridcount -2%
   \vfil\hhrule\vfil
   \llap{\smash{\raise -2.5pt
     \hbox{\L@Dc\number\gridcount\Zr@s\kern2pt}}}%
   \hhrule
   }

\def\vvrule{\vrule width \GL@wd \kern-\GL@wd}

 \def\vRule@{\advance\gridcount 2%
   \hfil\vvrule\hfil
   \setbox\auxGridbox@=\vbox to 0pt
      {\vskip \Htt@\vskip 2pt
        \hbox to 0pt{\hss\L@Dc\number\gridcount\Zr@s\hss}\vss}%
      \wd\auxGridbox@=0pt \box\auxGridbox@
   \vvrule
   }

 \def\PlaceGrid@@{\gridcount=10 
  \setbox\hGridbox@=\hbox{%
        \hbox{%
             \hskip-.4pt\vrule
             \vbox to \Htt@{%
               \offinterlineskip\parindent=\z@\relax
               \hbox to \Wdd@{\hfil}
               \hRule@\hRule@\hRule@\hRule@
               \vfil\hhrule\vfil}%
             \vrule\hskip-.4pt}
    }%
  \gridcount=0%
  \setbox\vGridbox@=\hbox{%
      \vbox{\offinterlineskip\parindent=0pt\hsize=0pt
         \vskip-.4pt\hrule%
         \hbox to \Wdd@{%
                 \vtop to \Htt@{\vfil}%
                 \vRule@\vRule@\vRule@\vRule@
                 \hfil\vvrule\hfil}%
         \hrule\vskip-.4pt}}%
  \wd\hGridbox@=0pt\ht\hGridbox@=0pt
  \wd\vGridbox@=0pt\ht\vGridbox@=0pt
  \hbox{\box\hGridbox@\box\vGridbox@}%
  }

 \def\LabelsGlobal{\def\LabGl@b{\global}}
 \def\LabelsLocal{\def\LabGl@b{}}
 \LabelsGlobal 

 \def\SetLabels#1\endSetLabels{%
   \LabGl@b\Labeltoks@={#1()\\}%
   }

 \LabGl@b\Labeltoks@={()\\}

 \def\ShowGrid{\LabGl@b\let\PlaceGrid@\PlaceGrid@@}
 \def\HideGrid{\LabGl@b\let\PlaceGrid@\relax}
 \def\Grids{\ShowGrid\LabGl@b\let\GridSwitch@\ShowGrid}
 \def\noGrids{\HideGrid\LabGl@b\let\GridSwitch@\HideGrid}

 \noGrids

 \def\bAdjust@@{%
     \setbox\auxLabelbox@=\hbox{\raise \dp\auxLabelbox@
            \box\auxLabelbox@}}
 \def\bAdjust@{\let\vAdjust@\bAdjust@@}

 \def\eAdjust@@{\dimen0=-.5\ht\auxLabelbox@
     \advance\dimen0 by .5\dp\auxLabelbox@
     \setbox\auxLabelbox@=
            \hbox{\raise\dimen0\box\auxLabelbox@}}
 \def\eAdjust@{\let\vAdjust@\eAdjust@@}

 \def\tAdjust@@{%
     \setbox\auxLabelbox@=\hbox{\raise-\ht\auxLabelbox@
            \box\auxLabelbox@}}
 \def\tAdjust@{\let\vAdjust@\tAdjust@@}

 \let\vAdjust@\relax

 \def\lAdjust@{\let\hAdjust@\rlap}
 \def\rAdjust@{\let\hAdjust@\llap}

 \let\hAdjust@\relax\let\vAdjust@\relax

 \def\FetchLabel@#1(#2)#3\\{%
     \IN@0#2@@\ifIN@
        \setbox0=\hbox{\ignorespaces#1#3\unskip}%
        \ifdim\wd0>0pt
           \ms@g{}%
           \ms@g{ !!! Bad label(s)? !!!}%
           \message{ #1(#2)#3}%
        \fi
        \def\LabelMole@##1\endFetchLabel@{%
            \IN@0()\\@##1@%
            \ifIN@\def\Temp@{\FetchLabel@##1\endFetchLabel@}%
            \else\def\Temp@{}%
            \fi
            \Temp@
           }%
     \else
       \ignorespaces#1\unskip
       \setbox\auxLabelbox@=%
         \hbox to 0pt{\hss\ignorespaces\hAdjust@
          {\ignorespaces#3\unskip}\hss}%
       \vAdjust@
       \let\hAdjust@\relax\let\vAdjust@\relax
       \AugmentLabelBox@@{#2}%
       \ht\Labelbox@=0pt\dp\Labelbox@=0pt
       \let\LabelMole@\FetchLabel@%
     \fi\LabelMole@}

 \newtoks\XYSep@ 
 \def\SetXYSeparator#1{%
     \IN@0#1@@\ifIN@\XYSep@{*}%
     \else
     \XYSep@{#1}%
     \fi
     }

 \SetXYSeparator*

 \def\AugmentLabelBox@@#1{%
     \IN@0\the\XYSep@ @#1@\ifIN@
       \SPLIT@0\the\XYSep@ @#1@%
       \setbox\Labelbox@=\hbox to 0pt{%
         \unhbox\Labelbox@
         \Shifted@@{\the\Initialtoks@\Wddd@}%
         {\the\Terminaltoks@\Httt@}%
         {\box\auxLabelbox@}}%
     \else
         \ms@g{}%
         \ms@g{ !!! Bad insertion point. !!!}%
         \message{ (#1\ this point was rejected.)}%
     \fi
    }

 \def\FetchOption@#1[#2]#3\endFetchOption@{%
    \def\temp{#1}
    \ifx\temp\empty
       \Edgect@=#2\relax
       \let\EdgGl@b\relax
       \SetEdges@
       \Cleaner@#3%
    \fi}

 \def\Cleaner@#1[@]{\Labeltoks@{#1}}
     
 \def\PlaceLabels@@{\mathsurround=0pt
     \def\Cr@{\\}%
     \let\L\lAdjust@\let\R\rAdjust@
     \let\B\bAdjust@\let\E\eAdjust@\let\T\tAdjust@
     \expandafter\FetchOption@\the\Labeltoks@[@]\endFetchOption@
     \Wddd@=\Wdd@ \Edgescale@\Wddd@ 
     \Httt@=\Htt@ \Edgescale@\Httt@
     \expandafter\FetchLabel@\the\Labeltoks@\endFetchLabel@
     \box\Labelbox@
     }%

 \let \PlaceLabels@\PlaceLabels@@

 \def\AffixLabels#1{\setbox\Coordinatebox@=\hbox{#1}%
      \Wdd@=\wd\Coordinatebox@ \Htt@=\ht\Coordinatebox@
      \advance\Htt@ \dp\Coordinatebox@
      \hbox{\copy\Coordinatebox@\kern-\Wdd@ 
           \Shifted@@{0pt}{-\dp\Coordinatebox@}%
           {\PlaceLabels@\PlaceGrid@}%
           \kern\Wdd@}%
      \GridSwitch@ 
      \LabGl@b\Labeltoks@{()\\}%
      }
 
   \let\wlog\LabelFigwlog@ld   
   \catcode`\@=\LabelFigCatAt  

                                By

              Raymond S\'eroul <A18645@FRCCSC21.BITNET>
                                and 
              Laurent Siebenmann <lcs@topo.math.u-psud.fr>
    
              VERSIONS: July 1991, Oct 1991, Jan 1992, July 1992

INTRODUCTION

      This labelling package is intended for TeX users who
rely on non-TeX sources for for their graphics inserts.  It
provides means for adding TeX labels to such inserts with a
minimum of fuss. 

       For most labels, TeX users have in the past found it
reasonably convenient to rely on non-TeX sources. Typical
occasions when an inescapable need for TeX labels seemed to
arise are

 (a) when the graphics program lacks certain exotic or complex
mathematical symbols

 (b) when the very highest typographical quality is wanted for the
labels

 (c) when labels included with the graphics fail to print, 
 and you cannot figure out why (cf. boxedeps.doc).  The labels
 provided by labelfig.tex are 100

       Since this package first appeared, many users, who in the
past scarcely dreamed of using TeX labels, have come to use
nothing but.  So it is now appropriate to add

Intoxication Warning:  TeX labels may be addictive and expensive. 

     If you have a fast preview you may disagree, and even find
that this package provides an agreeable paste-up environment; see
extra applications at end.

     Note to publishers: It is possible and convenient to ultimately
export the TeX labels produced by labelfig.tex to become an integral
part of the EPS file. This is often desired by a publisher who typically
uses an "upmarket" graphics or page layout program, with which the
staff is skilled in perfecting figures.  See Appendix I for
a recipe.

     The authors are grateful to Patrick Ion of Math Reviews for
helpful comments and encouragement.

BASIC INSTRUCTIONS

    After reading in the macro file using

preview or proof your figure with a coordinate grid printed on
top, by typing the following:

    \ShowGrid  
    \AffixLabels{<the graphics insertion>}

Here <the graphics insertion> is what you would type to insert
the graphics object alone without the grid.  This must provide
for the space around it. For example <the graphics insertion>
might well be \BoxedEPSF{MyFigure scaled 700} using the
boxedeps.tex macro package (from same source); this provides a
TeX box containing the encapsulated PostScript insert specified by
the file MyFigure. \AffixLabels{...} provides the grid (supposing
\ShowGrid is present) and later, once you have specified labels
using the grid, it will "tack on" the labels.

     The grid is a sort of (usually elongated) checkerboard of
ten rows and ten columns and its (internal) partitions are by
default numbered  .1, ... ,.9  both horizontally (X-coordinate
running left to right) and vertically (Y-coordinate running bottom
to top).  Thus the points enclosed by the grid correspond to the
points of the unit square in the cartesian "X-Y" plane, the lower
left corner corresponding to the origin (0,0).  By extrapolation,
the full page corresponds to a larger rectangle in the plane.

     These coordinates serve to position labels as follows.
Before the \AffixLabels{...} command type label specifications:

  \SetLabels
   (<X-coordinate>*<Y-coordinate>) <first label> \\
   .
   .
   .
   (<X-coordinate>*<Y-coordinate>)  <last label> \\
  \endSetLabels

Each row specifies one label and is terminated by \\.  In each
row, the position indicator comes first; it is written as a
standard cartesian point except that the X- and Y- coordinates
are separated by * rather than a comma because TeX allows a
comma as decimal point. There are no dimension units to specify
as the unit is the grid itself.

     By default, this cartesian point specifies where the middle
of the baseline of the label will be located.  However if you precede
the point by \L [or \R] the left [or right] edge of the baseline will
be located there. Similarly you may also precede the point by \T, \E,
or \B to vertically align the top equator or bottom of the label box
at the specified point.  This gives nine standard positions of
the label with respect to the insertion point --- corresponding to
the eight principle points of the compas and the center

                     \L\T     \T      \R\T

                     \L\E     \E      \R\E

                     \L\B     \B      \R\B

But this neglects the default "baseline" level of TeX,
giving potentially three more positions

                     \L    <no tag>   \R

For text, the baseline level is often the preferred. Its relation to
the others is variable. It will often coincide with the bottom level,
as happens for "X".  But it is often distinct, as for "g", in which
case you have in all 12 distinct positions rather than 9.

     It is convenient to think of this specification of label
position as attaching the label by a thumb-tack to the coordinate
grid. There are up to twelve positions of the thumb-tack on the
label, while the position of the thumb-tack on the coordinate grid is
arbitrary.  Normally, one choses the position of the thumb-tack on
the label to be the one that is the closest to the item being
labeled.  There are good reasons for this "rule of thumb":

   (a)  It facilitates correct positioning at first try.

   (b)  If the scale of the figure must be altered after labels
have been affixed, the labels have a good chance of remaining well
positioned.

   (c)  The visible grid need not extend beyond the "bounding box"
for the figure, because the best preferred position is always
(at least almost) within the bounding box .

The second reason is particularly important. Indeed it often
happens that scale has to be altered after labelling begins, in
order to either provide space for the labels, or to adjust
proportions between the labels and the figure.  (The size of labels
is unaffected by scaling.)

     Here is an artificial but self-contained test which uses
TeX rules to make a graphics object.

TEST

    Do not skip this!

 \def\FrameIt#1{\hbox{\vrule$\vcenter {\hrule\kern3pt%
             \hbox {\kern3pt #1\kern3pt}%
               \kern3pt\hrule}$\relax\vrule}}

 \def\Caption#1#2{\FrameIt{%
       \vtop {\hsize=#1\relax \parindent=0pt
         \leftskip=0pt \rightskip=0pt plus15pt
         \parfillskip=0pt
         \lineskip=1pt\baselineskip=0pt
         #2}}}

 \def\FirstQuadrant{\hbox to 100pt{\vrule\vbox to 100pt{%
        \hbox to 100pt{\hfil}\vfil\hrule}\hss}}

  \SetLabels
    \R(.5*.2) $\zeta\,\cdot$\\
    (.9*-.10) $\xi$\\
    \R(-.03*.9) $\eta$\\
    \T(.5*.9) \Caption{70pt}{%
          \it The norm of
          $g(\xi+i\eta)$ is indicated on
          contours of this invisible surface.}\\
  \endSetLabels

  \AffixLabels{\FirstQuadrant}

  \end

  Note that the coordinates to use for labels are indicated on the
edges of the grid (when visible) corresponding to the conventional
x- and y- axes of the Cartesian plane. By default the grid is
1-by-1. However, by the command \Edges{100}, you can change this
to 100-by-100 and many users find this alternative most
convenient. Place the command \Edges{...} in your style file (or
header) since its effect is is global. Other possible edge values
are 10 and 1000.

  If you use the command \Edges{...} at all, do so with care.  For
if you accidentally delete an \Edges{...} command your labels will
abruptly be badly misplaced and may logically but mysteriously
generate "dimension too big" errors under TeX and "off page" errors
under your driver.  

  You can dictate the edgescale for an individual figure by giving
the scale in brackets immediately after \SetLabels.  Thus, to
import into an article using say \Edge{100} a figure labelled using
another edgescale, say the original 1-by-1 default, you can use
\SetLabels[1]...\endSetLabels.

GETTING IT DOWN PAT

     Complicated labeling deserves the same respect as
complicated mathematics.  Do not expect it to come out perfect the
first time!  What is needed in either case is a mechanism to
repeatedly typeset troublesome pieces.

     One mechanism is always available.  One does complicated
labelling in a separate "test" file involving just the figure being
labelled;  a texpert will know how to \dump TeX's current state as
a temporary format that restarts rapidly at each retry.  Usually,
one then pastes the completed labelled figure back into the main
TeX file, but, of course, one can also \input it as an auxiliary
file.

     If you do not have a TeXpert at handy, here is a first
approximation to an efficient setup. By deletions reduce a copy
of your article to just a few lines before and after the figure.
Now label the figure, and finally, copy and paste the labelled
figure to the original article. Then copy the next figure to label
into this testbed and repeat. The TeXpert can improve the  speed
at which TeX starts up, by compiling a format specifically for
your article; just one caution: best NOT include in the format
ephemeral details of setup like \Set<mydriver>ArtSpecials (from
boxedeps.tex because this reads  figure dimensions which you may
change during your work session.

     An improved mechanism to repeatedly typeset troublesome
pieces is now available on the Macintosh; it is called LinoTeX;
see the same ftp sources.  It could be set up on many types
of computer.

     Before using labelfig.tex to attach labels to a graphics
object inserted using boxedeps.tex or BoxedArt.tex, make it a
firm rule to carefully adjust the bounding box using the trimming
commands of these packages, and also at least tentatively scale
and position the object. Beware of changing the grid inadvertently
after the labels have been positioned.  For example, correcting
the bounding box of a PostScript graphics object can foul up the
labels by changing the coordinate grid to which the labels are
attached. This is particularly true for the trimming  commands of
boxedeps.tex and BoxedArt.tex. However, as noted already, change
of scale is much less disruptive, and modest adjustments should be
well tolerated.

     Sometimes the labels protrude so far from the bounding box
of a figure that the figure has to be repositioned.  Best do this
by ad hoc spacing, say using \hglue and \vglue; altering the
bounding box would create a vicious circle.

     Remember that you are responsible for preventing labels
from overlapping. You are responsible for all label typography
including size and style. A label is really just about anything
that can be put in a TeX box. Note that spaces at the beginning
and end of labels will normally be suppressed; if you really want
them you must protect them with TeX braces.

     This package temporarily sets the \mathsurround parameter
of TeX to zero  while the labels are being affixed. This is done
because nonzero \mathsurround space would influence the position
of left and right aligned labels; then, when a texpert or printer
modifies mathsurround, diagram labeling might be disastrously
altered. There is a small price to pay involving labels that are
formatted as caption boxes including mathematics: you  may want or
need to specify an explicit mathsurround space within the caption
box; it will not influence anything outside.

     Those hostile to the use of * as separator between
the X and Y coordinates of label insertion points, are free to
impose another using \SetXYSeparator{<the new separator>}.  
Americans may prefer "," to "*" since they never use a 
comma as a decimal point; on the other hand, * may be more visible.

APPENDIX (I)  MERGING labelfig.tex LABELS INTO AN EPSF GRAPHICS OBJECT.

     As promised in the introduction, here is a recipe useful for
publishers. It works at least on Macintosh and at least for vectorized
graphics and Adobe type1 fonts.  (There is surely a similar recipe for
PCs under MSWindows.)

 (a)  Use boxedeps.tex utility to integrate the figure given by the eps
file, "x.eps" say, with a visible frame around it.  See
\ShowDisplacementBoxes command in boxedeps.tex.  To get precise results
automatically it is important to use the \Trim... commands of
boxedeps.tex making the "DisplacementBox" neatly fit the figure.

 (b)  Use the TeX printer driver and LaserWriter (versions >= 8.1.1) to
export to an EPSF the DVI page containing the integrated, labelled
figure. You now have an EPS file  "xx.eps"  that contains too much, and at
the wrong scale, and at wrong position.

 (c)  Convert the EPSF to an Adode Illustrator format EPSF using
the shareware utility called epsConvert by Sam Weiss
1993-- (currently $25).

 (d)  In Illustrator (or a compatible program), group the labels and the
"DisplacementBox"; copy them to the clipboard and paste them into "x.ps".
This step requires that all the label fonts be "visible to the Macintosh.

 (e)  Translate and scale the pasted group consisting of the labels plus
the "DisplacementBox" so as to make the "DisplacementBox" the bounding
box of (labelless) figure represented by "x.eps".  At this point the
labels will be correctly placed on the figure "x.eps".

 (f)  Ungroup and delete the "DisplacementBox".  The result is the
desired single EPS file, "x+.eps" say, It contains the original figure
plus its labels.  

     Using grouping and ungrouping appropriately in "x+.eps", a
publisher's staff can very efficiently improve label positions etc.

APPENDIX II)  SOME EXOTIC APPLICATIONS

     The grid of labelfig.tex is analogous to a light-table in
classical page makeup with wax or latex glue.  In principle, you
can use it to compose any page from its indivisible parts.  This
even has some of the artisanal charm of classical paste-up
provided you have a fast screen preview to make the process
"interactive".

     In practice labelfig.tex is a tool for nonstandard jobs.
Here are a few going beyond the labelling already discussed.

(I)  GRAPHICS INTEGRATION.

     This is accomplished by treating the imported graphics
objects as labels.  The underlying graphics object is then
typically an empty  \vbox to <dimension>{\vfill} in a TeX
\midinsert...\endinsert construction.  A label line
might be of the form

   (.1*.1) \\

The exact form of the special command varies from driver to
driver.  However, in the case of encapsulated PostScript graphics
(EPSF norm), by relying on boxedeps.tex, one can have the
following standard syntax (independant of driver  (see
boxedeps.doc for details.
  
  (.1*.1) \BoxedEPSF{MyFigure scaled <scale in mils>}\\

This may be slow since it requires TeX to read the PostScript
file to read bounding box using many complex macros.  So you
may want to try

  (.1*.1) \EPSFSpecial{MyFigure}{<scale in mils>}\\

which is fast and driver independant, but it squashes the
bounding box, normally to its lower left corner.

     Similarly for graphics of the Macintosh PICT norm ---
using BoxedArt.tex (same sources) in place of boxedeps.tex.

     This approach to integration is to be recommended when
one is assembling a composite graphics object.

 (II)  COMMUTATIVE DIAGRAM ENHANCEMENT

     Commutative diagrams or arrays of mathematical objects
connected by arrows of various sorts are common in mathematics.
The mathematical objects require the use of TeX.  Recently TeX
acquired a good collection of arrows of all slopes --- that of
LamSTeX --- plus pwerful macros to build the diagrams.

     However, even the LamSTeX collection is often
inadequate; it lacks for example double shafted arrows, dotted
arrows and curved arrows. Fortunately it is possible to produce
such arrows on an individual basis using sophisticated graphics
programs such as Illustrator and AldusFreehand (both serving
the EPSF norm) or using Metafont (with its public domain norm).
Since the creation of each new arrow is a work of love, you
probably want to limit the number of arrows by using LamSTeX
for most arrows. The 40K commutative diagram module of LamSTeX
has been adapted to work with AmSTeX and a copy may be posted
with LabelFig and related files. Unfortunately no one has yet
offered a version that works with Plain TeX or LaTeX.

       Suffice it here to say that when the exotic arrow has
been somehow imported into TeX, labelfig.tex treats it as a
label that one affixes to the commutative diagram.  Two other
steps will be treated in separate notes, namely the matter of
extracting the dimension specifications for the arrow and the
construction of the arrow --- for these steps are far from
unique and often depend intimately on your computer environment. 
Notes for the Macintosh-Textures-Illustrator combination are
found in the file ExoticArrows.doc.

 (III) NESTING 

Ingenuity pays off in exploiting labelfig.tex. One can
mix graphics and typography quite freely.  labelfig.tex is good
for freeform or overlapping arrangements, while boxedeps.tex (or
BoxedArt.tex) is best for regimented non-overlapping
arrangements --- and the two can be combined.

     The default behavior of labelfig.tex is not ideal 
for nesting objects, because to prevent trouble for beginners
the register for labels is globally cleared when \AffixLabels
concludes.  But there are switches available

      \LabelsGlobal      \LabelsLocal

which change this.  To understand this, extend the above test 
by something like:

 \LabelsLocal

 \SetLabels
    (.5*.5) AAA\\
 \endSetLabels

 {
 \SetLabels
    (.5*.5) ZZZ\\
 \endSetLabels
   \AffixLabels{\FirstQuadrant}
 }

   \AffixLabels{\FirstQuadrant}

     There are however potential pitfalls.  Neither
labelfig.tex nor boxedeps.tex has been tested under extreme
conditions. Problems may occur if their procedures are
indiscriminately nested. For boxedeps.tex (not labelfig.tex)
there is a precise cause for worry, namely many of its
variables are "global", which means that TeX braces will not
provide the protection one might expect.

COMMAND SUMMARY FOR labelfig.tex

  Here [...] means optional (one or zero)
       [...]* means any number of such constructs

  \SetLabels
    [[<P>](<X><Sep><Y>) <label> \\]*
  \endSetLabels
  \ShowGrid  
  \AffixLabels{<the figure>}

   --- <P> is tack position, one of eleven or empty
              order irrelevant

                   \L\T      \T      \R\T

                   \L\E      \E      \R\E

                     \L               \R

                   \L\B      \B      \R\B

   --- (<X><Sep><Y>) insertion point;
  <Sep> is separator, = * by default;
  \SetXYSeparator{<Sep>} changes it.
   <X> and <Y> are real numbers

  --- <label> a label to attach 

  --- <the figure> the figure to label 

  \GlobalLabels (default)     
  \LocalLabels  setting for nested constructs.

 \Grids makes ALL grids appear; \HideGrid then makes just next disappear.
 \noGrids returns to default.  The commands are always global.

 \GridLineWidth{<dimension>} adjusts width of grid lines. Default is very
small, to give "hairline" effect. If your grid lines are missing try
setting \GridLineWidth{1pt}.

 \Edges#1 globally changes the edge size of all grids to the numerical 
value #1, which must be 1, 10, 100, or 1000.  The default is 1.

VERSION HISTORY.
 --- Jan 1993: \Edges#1 and [??] option after \SetLabels
 --- July 1992: \Grids, \noGrids, \HideGrid;
       Gridlines become hairlines; \GridLineWidth{<dimension>}.
 --- Oct 1991, Jan 1992: \SetXYSeparator{<Sep>},  \LabelsGlobal,
       \LabelsLocal.
 --- July 1991: first release

Address for bugs and other feedback:

        Raymond S\'eroul
        IREM and Lab. de Typographie Informatise
        Univ. Rene Descartes
        Strasbourg

    Tel 33-88-41-63-45
    Email:  A18645@FRCCSC21.BITNET

        Laurent Siebenmann
        Mathematique, Bat. 425,
        Univ de Paris-Sud,
        91405-Orsay,
        France

    Tel 33-1-6941-7949; 
    Email: lcs@topo.math.u-psud.fr

\def\scalefig#1{\epsfxsize #1\textwidth}

\def\bs{\boldsymbol}

\newcommand {\Ebb}{{\mathbb{E}}}

\newcommand{\tr}{\mbox{${\mbox{tr}}$}}

\newtheorem{theorem}{Theorem}
\newtheorem{lemma}{Lemma}

\newtheorem{algorithm}{Algorithm}

\title{\LARGE {Filter-and-Forward Transparent Relay Design for OFDM Systems}}

\author{
Donggun Kim, Junyeong Seo, {\em Student~Members, IEEE}, and
\\Youngchul Sung$^\dagger$\thanks{$^\dagger$Corresponding author},
{\em Senior~Member, IEEE}
\thanks{The authors are with the Dept. of Electrical Engineering,  KAIST, Daejeon 305-701, South
Korea. E-mail:\{dg.kim@, jyseo@, and ysung@ee.\}kaist.ac.kr. 
This research was also supported  by the KCC (Korea
Communications Commission), Korea, under the R\&D program
supervised by the KCA (Korea Communications Agency)
(KCA-2012-11-911-04-001).  Some part of the paper was presented  in \cite{DGKim:12APSIPA}.}
}

\begin{document}

\maketitle

\begin{abstract}
 In this paper, the filter-and-forward (FF) relay design for
orthogonal frequency-division multiplexing (OFDM) transmission
systems is considered to improve the system performance over
simple amplify-and-forward (AF) relaying. Unlike conventional OFDM
relays performing OFDM demodulation and remodulation, to reduce
processing complexity, the proposed FF relay directly filters the
incoming signal in time domain with a finite impulse response (FIR) and forwards
the filtered signal to the destination. Three design criteria are
considered to optimize the relay filter. The first criterion is the minimization of the relay transmit power subject to
per-subcarrier signal-to-noise ratio (SNR) constraints, the second is the maximization of the worst subcarrier channel SNR subject to
source and relay transmit power constraints, and the third is the maximization of data rate  subject to
source and relay transmit power constraints.  It is shown that the
first problem reduces to a semi-definite programming (SDP) problem by semi-definite relaxation and the solution to the relaxed SDP problem has rank one under a mild condition.  For the latter two problems,  the problem of joint source power allocation and
relay filter design is considered  and an efficient algorithm is proposed for each problem based on alternating optimization and the projected gradient method (PGM).  Numerical results show that the
proposed FF relay significantly outperforms simple AF relays with insignificant increase in complexity. Thus, the proposed FF relay provides a practical
alternative to the AF relaying scheme for OFDM transmission.
\end{abstract}


\section{Introduction}

\vspace{-0.5em}

Recently, relay networks have drawn extensive interest from the
research community because they play an important role in
enlarging the network coverage and improving the system
performance in current and future wireless networks. Indeed,
LTE-Advanced adopts relays for coverage extension and performance
improvement \cite{Hoymannetal:12ComMag}. There are several
well-known relaying schemes such as AF, decode-and-forward (DF),
and compress-and-forward (CF)
\cite{Cover&ElGamal:79IT,ElGamal&Aref:82IT,Kramer:05IT}. Among the
relaying schemes, the AF scheme (i.e., simple repeater) is the simplest and is suitable
for cheap relay deployment under transparent\footnote{
Transparent operation means that the destination node does not
know the existence of the relay node, and this operation is
suitable for cheap AF relays \cite{Hoymannetal:12ComMag}.}
operation \cite{Hoymannetal:12ComMag}. Recently, there have been
some efforts to extend this simple AF scheme to a linear filtering
relaying scheme, i.e., an FF scheme, to obtain better performance
than the AF scheme while keeping the benefit of low computational
complexity of the AF scheme \cite{ElGamal&Mohseni&Zahedi:06IT, DelCoso&Ibars:09WC,
Chen10:SP,Liang:11WCOM, SungKim:11arXiv, KimSungLee:12SP}. It has been shown that the FF scheme can
outperform the AF scheme considerably. However, most of the
previous works on the FF relay have been done for single-carrier
transmission, whereas most of the current wireless standards adopt OFDM
transmission. Thus, in this paper, we propose direct FF relaying
for OFDM transmission instead of using conventional OFDM relays
which OFDM-demodulate the incoming signal, amplify or decode the
demodulated signal, OFDM-remodulate the processed signal and
transmit the remodulated OFDM signal to the destination
\cite{Hammerstr:06ConfCommun,Ng:07JSAC,Dong:10ICASSP,Dang:10WC}.
In the proposed scheme, the incoming signal to the relay is FIR
filtered at the {\it chip rate} of the OFDM modulations in time domain and the
filtered signal is directly forwarded to the destination. In this
way, the necessity of OFDM processing at the relay is eliminated,
but the overall performance can still be improved over the AF
scheme by a properly designed relay filter.

\begin{figure}[htbp]
\centerline{
    \begin{psfrags}
    \psfrag{RelayCell}[c]{Relay cell} %
    \psfrag{Relay}[c]{Relay} %
    \psfrag{BS1}[c]{{\small Basestation}} %
    \psfrag{MS1}[c]{{\small Terminal station}} %
    \scalefig{0.5}\epsfbox{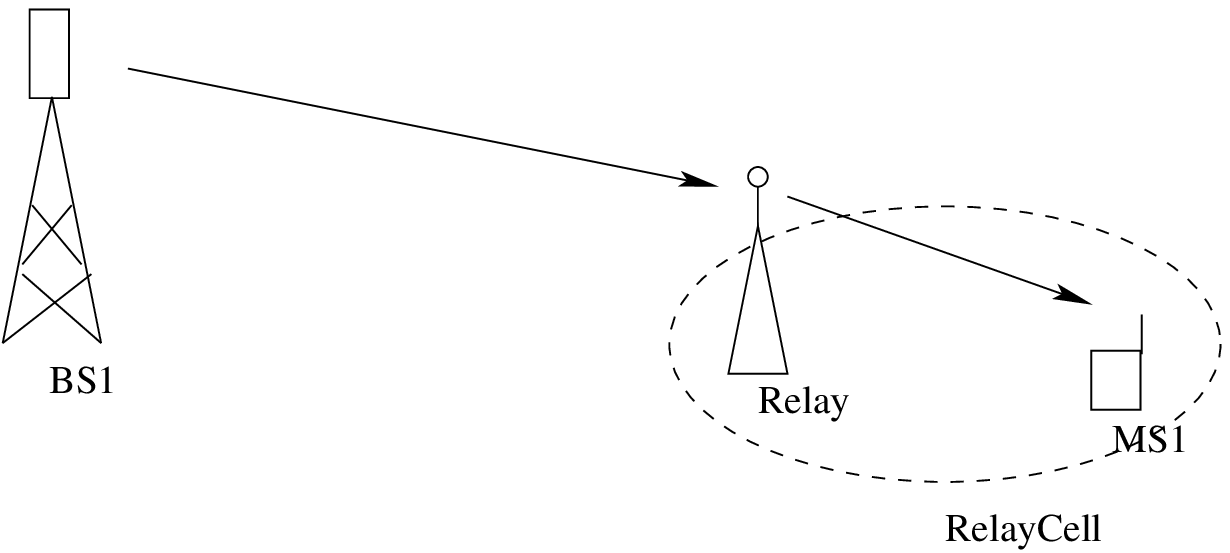}
    \end{psfrags}
} \caption{The considered relay network} \label{fig:systemModel}
\end{figure}

\subsection{Our Approach and Contributions}

{
In the paper, we consider three meaningful criteria for the FF relay design for OFDM
systems, minimization of power consumption at the relay,  maximization of  the worst subcarrier SNR, and maximization of the data rate,
under the scenario of  single-input single-output (SISO)
communication as the first step in this research
direction. (The multiple-input multiple-output (MIMO)
communication case is beyond the scope of this paper and will be
studied as a future work.) Our contributions on this topic are summarized in the below.
\begin{itemize}
\item First, exploiting the eigen-property of circulant matrices  and the structure of
Toeplitz filtering matrices, we derived necessary expressions for the problem formulation such as the subcarrier SNR and the relay transmit power in terms of the design variables of the relay filter coefficients and source power allocation, and formulated the above problems explicitly based on the derived expressions. %
\item In the case of the first problem of  relay power minimization under per-subcarrier SNR constraints, we showed that the problem is expressed as a semi-definite relaxation  problem. That is, the original non-convex FF relay design
problem is  approximated by a convex SDP problem. Furthermore, in this case  we  showed that the solution to the relaxed SDP
problem is the same as that to the original problem under a 
mild condition. %
\item For the second design criterion, we formulated the problem of joint design of the relay filter and the source power allocation for 
the worst subcarrier SNR maximization subject to source and  relay power constraints and provided an efficient iterative algorithm to solve this problem based on alternating optimization. The provided algorithm consists of two steps at each iteration, optimizing the relay filter to maximize the worst subcarrier SNR for given power allocation and optimizing the source power allocation to maximize the worst subcarrier SNR for a given relay filter, and guarantees convergence to a locally optimal point, although the convergence to a globally optimal point is not guaranteed.  We  showed that the first step of the iteration reduces to a SDP problem and the second step of the iteration reduces to a linear programming (LP).    The second criterion is closely related to bit error rate (BER) minimization in case of weak or no channel coding in addition to 
  overall quality-of-service (QoS) improvement for subcarrier channels.  This is because   in the single-user case, bits for one user are distributed across subcarriers and the overall system BER is dominated by the BER of the worst subcarrier channel \cite{Dembo&Zeitouni:book}. %
\item For the third problem of joint optimization of the relay filter and the source power allocation for rate maximization, we proposed an efficient algorithm by applying  the projected gradient method \cite{goldstein64, Polyak:69USSR, Slavakis&Yamada&Ogura:06NFAO, KimSungLee:12SP} directly to this constrained optimization problem. The proposed method guarantees the satisfaction of the constraints and convergence to a locally optimal point. %
\end{itemize}
}

Numerical results
show that the proposed FF relay significantly outperforms simple
AF relays  and furthermore achieves most of the performance gain with not so many filter taps. Thus, the proposed FF relay
provides a practical alternative with low complexity to the AF
relaying scheme for OFDM transmission.

\subsection{Notation and Organization}

In this paper, we will make use of standard notational
conventions. Vectors and matrices are written in boldface with
matrices in capitals. All vectors are column vectors.  For a matrix $\Abf$,
$\Abf^*$, $\Abf^T$, $\Abf^H$, and $\mbox{tr}(\Abf)$ indicate the
complex conjugate, transpose, conjugate transpose, and trace of
$\Abf$, respectively.  $\Abf \succeq 0$ and $\Abf \succ 0$ mean
that $\Abf$ is positive semi-definite and that $\Abf$ is strictly
positive definite, respectively. For two matrices $\Abf$ and $\Bbf$, $\Abf
\succeq \Bbf$ means that $\Abf-\Bbf \succeq 0$. $\Ibf_n$ stands
for the identity matrix of size $n$ (the subscript is omitted
    when unnecessary), and ${\mathbf {0}}_{m \times n}$ denotes a $m
\times n$ matrix  with all zero elements. The notation
${\mathtt{Toeplitz}}(\fbf^T,N)$ indicates a $N \times (N+L_f-1)$
Toeplitz matrix with $N$ rows  and  $[ \fbf^T, 0, \cdots, 0 ] $ as
its first row vector, where $\fbf^T$ is a row vector of size
$L_f$, and $\mbox{diag}(d_1,\cdots,d_n)$ means a diagonal matrix with
diagonal elements $d_1,\cdots,d_n$. The notation $\xbf\sim
{\cal{CN}}(\mubf,\Sigmabf)$ means that $\xbf$ is complex
circularly-symmetric Gaussian distributed with mean vector $\mubf$
and covariance matrix $\Sigmabf$. $\Ebb\{\cdot\}$ denotes the
expectation. $j=\sqrt{-1}$.

The remainder of this paper is organized as follows. The system model is described in
Section \ref{sec:systemmodel}. In Section \ref{sec:FFrelayProblem},
the FF relay design problems are formulated and solved. In Section \ref{sec:NumericalResult}, the performance of the proposed design
methods  is investigated. {Several issues regarding practical implementation of the proposed FF relay are discussed in Section \ref{sec:discussion},} 
followed by the conclusion in Section
\ref{sec:conclusion}.

\section{System Model}
\label{sec:systemmodel}

\begin{figure}[htbp]
\centerline{
   \begin{psfrags}
\psfrag{s1}[c]{{\footnotesize $s[k]$}} %
\psfrag{s2}[c]{{\footnotesize $s[N-1]$}} %
\psfrag{s3}[c]{{\footnotesize $s[0]$}} %
\psfrag{s5}[c]{{\footnotesize $x[N-1]$}} %
\psfrag{vd}[c]{{\footnotesize $\vdots$}} %
\psfrag{s6}[c]{{\footnotesize $x[N-L]$}} %
\psfrag{s7}[c]{{\footnotesize $x[0]$}} %
\psfrag{sp}[c]{{\footnotesize S/P}} %
\psfrag{ps}[c]{{\footnotesize P/S}} %
\psfrag{idft}[c]{{\footnotesize IDFT}} %
\psfrag{dft}[c]{{\footnotesize DFT}} %
\psfrag{hsr}[c]{{\footnotesize $f[l]$}} %
\psfrag{hrd}[c]{{\footnotesize $g[l]$}} %
\psfrag{hff}[c]{{\footnotesize $r[l]$}} %
\psfrag{trans}[c]{{\footnotesize Source}} %
\psfrag{FFrelay}[c]{{\footnotesize FF relay}} %
\psfrag{x}[c]{{\footnotesize $\tilde{x}_s[n]$}} %
\psfrag{ToDestination}[l]{{\footnotesize Destination}} %
\scalefig{1}%
\epsfbox{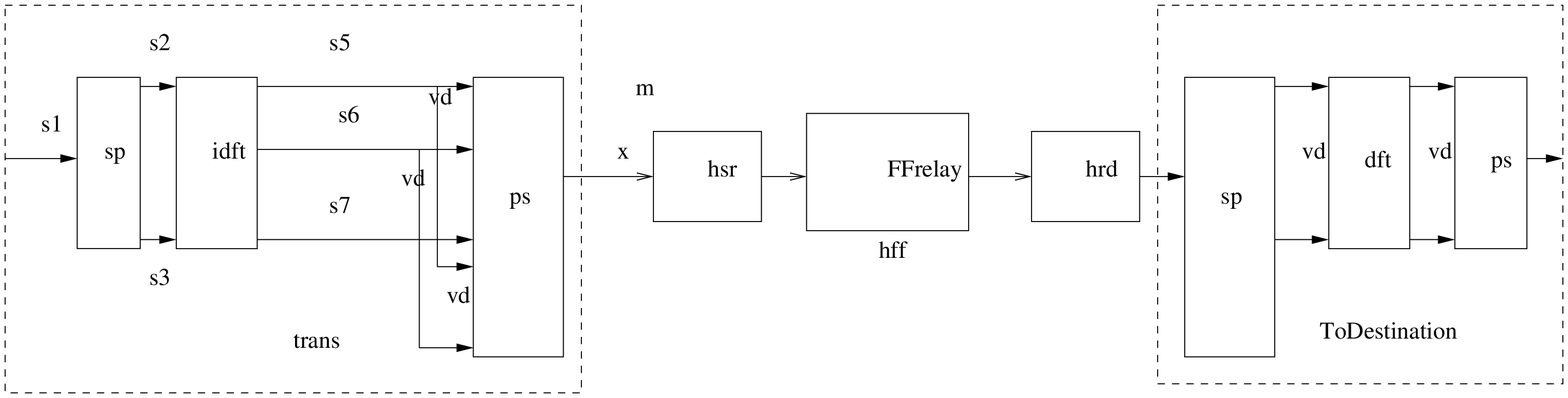}%
\end{psfrags}
} \caption{The considered relay network with an OFDM transmitter,
an FF relay, and a destination node} \label{fig:systemModel2}
\end{figure}

We consider a full-duplex\footnote{Please see Section \ref{sec:discussion}.} relay network composed of a source node
(basestation), a relay node and a destination node (terminal
station), as shown in Figures  \ref{fig:systemModel} and
\ref{fig:systemModel2}, where the source employs OFDM modulation
with $N$ subcarriers and  each link performs SISO communication. We
consider the case that the direct link between the source and the
destination is {seriously faded}. Thus, for simplicity, we assume that there is
no direct link between the source and the destination and that
both the source-to-relay (SR) link and the
relay-to-destination (RD) link are {frequency-selective 
channels modeled as multi-tap filters with finite impulse responses.} We assume that the relay is an FF
relay, i.e., the relay performs FIR filtering on the incoming
signal at the chip rate of the OFDM modulation and transmits the
filtered output immediately to the destination.\footnote{Such an FF scheme requires up and down converters and simple baseband circuitry  only.} Thus, the FF relay
can be regarded as {an additional frequency-selective (time-dispersive) channel} between the source
and the destination. We assume that the order of the FIR filter
at the relay does not make the length of the overall FIR
channel between the source and the destination larger than that of
the OFDM cyclic prefix.  Since our focus of using the FF relay in this paper
is the transparent relay operation, we assume that the SR channel
state is known to the relay and that the RD channel state is
unknown to the relay but the RD channel distribution is known to
the relay. Such an assumption is reasonable for the
transparent relay operation since the relay node does not have its own identity and the destination node does not know
the existence of the relay; thus, the destination node does not
feedback information to the relay node directly. (Please see Section \ref{sec:discussion} regarding how to obtain channel information.)

Specifically,  at the source, the length $N$ data vector of OFDM
symbols is given by
  $\sbf := [ s[N-1], s[N-2], \cdots ,$ $s[0] ]^T$,
where each data symbol is assumed to be a zero-mean independent
complex Gaussian random variable with variance $P_{s,k}$, i.e.,  $s[k] \sim {\mathcal{CN}}
(0, P_{s,k})$ for $k=0,1,\cdots,N-1$. The time-domain signal vector
$\xbf_s$ after normalized inverse discrete Fourier transform (IDFT)
at the source is given by
\begin{equation} \label{eq:timeseq} \underbrace{\left [
\begin{array}{c}
{x_s[N-1]} \\
{x_s[N-2]} \\
\vdots \\
{x_s[0]}
\end{array}
 \right ]
 }_{=:\xbf_s}
 =  \underbrace {\frac{1}{\sqrt{N}}\left [
 \begin{array}{cccc}
 1 & 1     & \cdots  & 1 \\
 1 & \omega_N     & \cdots  & \omega_N^{N-1} \\
 \vdots &  \vdots & \ddots & \vdots \\
 1 & \omega_N^{N-1}  & \cdots & \omega_N^{(N-1)^2}
 \end{array}
 \right ]}_{=:\Wbf_N}
\underbrace{ \left [
\begin{array}{c}
{s[N-1]} \\
{s[N-2]} \\
\vdots \\
{s[0]}
\end{array}
 \right ]}_{=\sbf},
\end{equation}
where $\omega_N = e^{j \frac{2 \pi}{N}} $. Let $\wbf_{k-1}^T$
denote the $k$-th row of the normalized IDFT matrix  $\Wbf_N$ for
$k = 1, \cdots, N$. Then,  $\xbf_s$ can be written  as
 $\xbf_s = [\wbf_0^T \sbf,~~ \wbf_1^T \sbf,~ \cdots, ~~\wbf_{N-1}^T\sbf
 ]^T$
 and the covariance matrix $\Sigmabf_{\xbf_s}$ of $\xbf_s$ is given
by
\begin{equation} \label{eq:covarianceMat}
 \Sigmabf_{\xbf_s} = \Ebb\{\xbf_s
{\xbf_s}^H \}= \left[
\begin{array}{cccc}
\frac{1}{N}\sum_{k=0}^{N-1} P_{s,k} & {\frac{1}{\sqrt{N}}\pbf_s^T \wbf_1^*} & \cdots & {\frac{1}{\sqrt{N}}\pbf_s^T \wbf_{N-1}^*} \\
{\frac{1}{\sqrt{N}}\wbf_1^T \pbf_s} & \frac{1}{N}\sum_{k=0}^{N-1} P_{s,k}  & \cdots & {\frac{1}{\sqrt{N}}\pbf_s^T \wbf_{N-2}^*} \\
\vdots & \vdots  & \ddots & \vdots \\
{\frac{1}{\sqrt{N}}\wbf_{N-1}^T \pbf_s} & {\frac{1}{\sqrt{N}}\wbf_{N-2}^T \pbf_s}  & \cdots
&\frac{1}{N}\sum_{k=0}^{N-1} P_{s,k} \\
\end{array}
\right ],
\end{equation}
where  $\pbf_s = [P_{s,N-1}, P_{s,N-2}, \cdots , P_{s,0}]^T$ is {the vector composed of source power assigned to each subcarrier}, since
\begin{equation}  \label{eq:CovTildeX} \Ebb\{\wbf_i^T\sbf \sbf^H
\wbf_j^* \} = \left\{
\begin{array}{ll}
{\frac{1}{\sqrt{N}}\wbf_{i-j}^T\pbf_s} & \mbox{if}~ i > j,\\
 \frac{1}{N}\sum_{k=0}^{N-1} P_{s,k} & \mbox{if}~ i = j, \\
{\frac{1}{\sqrt{N}}\pbf_s^T \wbf_{j-i}^*} & \mbox{if}~ i < j.
\end{array}
\right.
\end{equation}
 The vector $\xbf_s$ is attached by a cyclic prefix with length
$L_{CP}$, i.e.,
\begin{equation}
\tilde{x}_s[n] = \left \{
\begin{array}{ll}
x_s[n],  &~~~~~~ n = 0, 1, \cdots, N-1,\\
x_s[N+n], &~~~~~~ n = -1, -2, \cdots, -L_{CP},
\end{array}
\right.
\end{equation}
 and the cyclic prefix attached sequence $\tilde{x}_s[n]$
 is transmitted from the source to the relay through the SR
 channel. Then, the received baseband signal at the relay is given by
\begin{equation}
y_r[n] = \sum_{l=0}^{L_f-1}f_{l}\tilde{x}_s[n-l] + n_r[n],
\end{equation}
where $\fbf=[f_0,f_1,\cdots,f_{L_f-1}]^T$ is the channel tap
coefficient vector of the SR channel known to the relay, $L_f$ is
the length of the SR FIR channel, and $n_r[n]$ is the additive
white Gaussian noise at the relay with  $n_r[n] \sim \Cc\Nc(0,
\sigma_r^2)$. At the relay, the received signal $y_r[n]$ is {\it
FIR filtered at the chip rate of the OFDM transmission} and then
transmitted immediately to the destination. Thus, the output
signal at the relay at (chip) time $n$ is given by
\begin{equation} \label{relay:transmitted}
y_t [n] = \sum^{L_r -1}_{l=0} r_l ~ y_r[n-l],
\end{equation}
where $\rbf = [r_0,r_1,\cdots,r_{L_r-1}]^T$ is the FIR filter
coefficient vector at the relay and $L_r$ is the order of the FIR
filter.  Note that, when $L_r=1$, the FF relay simply reduces to
the AF relay. However, when $L_r > 1$, the FF relay is an
extension of the AF relay with some amount of digital processing.
Finally, the signal transmitted by the relay goes through the RD
FIR channel to the destination.  Thus, the received signal at the
destination is given by
\begin{equation}
y_d [n] =\sum^{L_g -1}_{l=0} g_l ~ y_t[n-l] + n_d[n],
\end{equation}
where $\gbf = [g_0,g_1,\cdots, g_{L_g-1}]^T$ is the FIR channel tap
coefficient vector for the RD channel,  $L_g$ is the order of the RD
FIR channel, and  $n_d[n]$  is zero-mean white Gaussian noise with
variance $\sigma_d^2$  at the destination. Here, we assume that the
channel tap coefficient $g_l$, $l=0,1,\cdots,L_g-1$, is independent
and identically distributed (i.i.d.) according to $g_l
\stackrel{i.i.d.}{\sim} \Cc\Nc(0, \sigma_g^2)$, i.e., each tap is
independently Rayleigh faded, and that the realization $\{g_l,l=0,1,\cdots,L_g-1\}$ is not known to the relay but its distribution is known to the relay.  By stacking the output symbols
at the  relay and the received symbols at the destination, we have
the following vectors for the transmitted signal at the relay and
the cyclic prefix portion removed received signal vector  at the
destination, respectively:
\begin{equation} \label{eq:RelayTransmittedSignal}
\ybf_t =  \Rbf\Fbf\tilde{\xbf}_s + \Rbf \nbf_r ~~~\mbox{and}~~~
\ybf_d =  \Gbf\Rbf\Fbf\tilde{\xbf}_s + \Gbf\Rbf\nbf_r +\nbf_d,
\end{equation}
where
\begin{eqnarray}
\ybf_d &=& \left [{y_d}[N-1],  {y_d}[N-2], \cdots, {y_d}[0]
\right]^T, \nonumber \\
\ybf_t &=& \left [{y_t}[N-1], {y_t}[N-2], \cdots, {y_t}[0],
y_t[-1], \cdots ,{y_t}[-L_g+1] \right]^T, \nonumber \\
\tilde{\xbf}_s &=& [ x_s[N-1], x_s[N-2], \cdots, x_s[0], x_s[-1],
\cdots,
x_s[-L_g-L_r-L_f+3]]^T, \label{eq:xs_tilde} \\
\nbf_r &=& \left [ n_r[N-1],  n_r[N-2], \cdots,  n_r[0], n_r[-1]
\cdots, n_r[-L_g-L_r+2] \right]^T,\nonumber \\
\nbf_d &=& \left [ n_d[N-1], n_d[N-2], \cdots,  n_d[0]
\right]^T,\nonumber\\
\Gbf &=& {\mathtt{Toeplitz}}(\gbf^T,N),\nonumber\\
\Rbf &=& {\mathtt{Toeplitz}}(\rbf^T,N+L_g-1),\nonumber\\
\Fbf &=& {\mathtt{Toeplitz}}(\fbf^T,N+L_g+L_r-2). \nonumber
\end{eqnarray}
Under the assumption that $L_{CP} \ge L_g+L_r+L_f -3$, the DFT of
the cyclic prefix portion removed received vector of size $N$ at the
destination is given by
\begin{eqnarray}
\hat{\ybf}_d &=& \Wbf_N^H\Gbf\Rbf\Fbf\tilde{\xbf}_s + \Wbf_N^H\Gbf\Rbf\nbf_r +\Wbf_N^H\nbf_d, \\
             &=& \Wbf_N^H{\Hbf_c}\Wbf_N\sbf + \Wbf_N^H\Gbf\Rbf\nbf_r +\Wbf_N^H\nbf_d, \label{eq:FinalYdinSecIIAm1}\\
             &=& \Dbf\sbf + \Wbf_N^H\Gbf\Rbf\nbf_r +\Wbf_N^H\nbf_d,
             \label{eq:FinalYdinSecIIA}
\end{eqnarray}
where $\Wbf_N^H$  is the normalized DFT matrix of size $N$,
$\Hbf_c$ is a $N\times N$ circulant matrix generated from the
overall Toeplitz filtering matrix $\Gbf\Rbf\Fbf$ from the source
to the destination, and
$\Dbf=\mbox{diag}(d_0,\cdots,d_{N-1})=\Wbf_N^H{\Hbf_c}\Wbf_N$ is
the eigen-decomposition of $\Hbf_c$.

\subsection{Manipulation for quadratic forms}

The received signal form \eqref{eq:FinalYdinSecIIA} is standard in
OFDM transmission, but the form cannot be  used directly for relay
filter optimization in the next section. Thus, in this subsection, we
derive an explicit expression for the received signal
$\hat{y}_d[k]$, $k=0,1,\cdots,N-1$, which facilitates optimization
formulation in the next section, based on the following property
of circulant matrices \cite{Gray:06Toeplitzbook}.

\vspace{0.5em}
\begin{lemma} \label{lem:circulantM} \cite{Gray:06Toeplitzbook}
 Let $\Cbf$ be an $N\times N$ circulant matrix with the first row
$[c(0), c(1), \cdots, c(N-1)]~$. Then, the eigenvalues of $\Cbf$ are
given by
\[
\lambda_k = \sum_{n=0}^{N-1} c(n) \omega_N^{-kn}, ~~~k = 0,1,
\cdots,N-1,
\]
with the corresponding right eigenvectors
\[
\xibf_k = \frac{1}{\sqrt{N}} [ 1, \omega_N^{-k}, \omega_N^{-2k},
\cdots, \omega_N^{-(N-1)k}]^T,~~~k = 0,1, \cdots, N-1.
\]
\end{lemma}
\vspace{0.5em}

By Lemma \ref{lem:circulantM}, to derive the diagonal
elements of  $\Dbf$ in \eqref{eq:FinalYdinSecIIA}, we need to know
only the first row of $\Hbf_c$ in \eqref{eq:FinalYdinSecIIAm1}.
Let the first row of $\Gbf$ be $\tilde{\gbf}^T$. Then,
 $\tilde{\gbf}^T$ is a $1 \times (N+L_g-1)$ row vector given by
\begin{equation} \label{eq:gbftilde}
\tilde{\gbf}^T=[\gbf^T,0,\cdots,0],
\end{equation}
and the first row of $\Gbf\Rbf\Fbf$ is given by $\tilde{\gbf}^T
\Rbf \Fbf$. Since $\Hbf_c$ is generated by truncating out the
elements of $\Gbf\Rbf\Fbf$ outside the first $N \times N$
positions and by moving the lower $(L_g+L_r+L_f-3) \times
(L_g+L_r+L_f-3)$ elements of the truncated part to the lower left
of the untruncated $N \times N$ matrix, the first row $\hbf_c^T$
of $\Hbf_c$ is simply the first $N$ elements of the first row
$\tilde{\gbf}^T \Rbf \Fbf$ of $\Gbf\Rbf\Fbf$, i.e., $\hbf_c^T =
\tilde{\gbf}^T\Rbf\Fbf\Tbf$, where $\Tbf$ is a truncation matrix
for truncating out the elements of $\tilde{\gbf}^T\Rbf\Fbf$ except
the first $N$ elements, given by
\begin{equation}
\Tbf = \left [
\begin{array}{l}
\mbox{I}_N \\
{\bf{0}}_{(L_g+L_r+L_f-3) \times N}
\end{array} \right].
\end{equation}
 Now, the diagonal elements of $\Dbf$ can be obtained by Lemma  \ref{lem:circulantM} and are given by
\begin{equation}
[d_0,\cdots,d_{N-1}]^T =  \sqrt{N} \Wbf_N^H
(\tilde{\gbf}^T\Rbf\Fbf\Tbf)^T,
\end{equation}
where  $\sqrt{N}\Wbf_N^H$ is the DFT matrix of size $N$. Finally,
the received signal in the $k$-th subcarrier at the destination
is expressed as
\begin{equation} \label{eq:ReceivedSig_perCarrier}
\hat{y}_d[k] =  \sqrt{N}\wbf_k^H \Tbf^T \Fbf^T \Rbf^T \tilde{\gbf}
s[k] + \wbf_k^H\Gbf\Rbf\nbf_r +\wbf_k^H\nbf_d,
\end{equation}
where $\wbf_k^H$ is the ($k+1$)-th row of $\Wbf_N^H$. Thus, the
signal and noise parts of $\hat{y}_d[k]$ are  given by
\begin{equation} \label{eq:ReceivedSignal}
\hat{y}_{d,S}[k] =  \sqrt{N}\wbf_k^H \Tbf^T \Fbf^T \Rbf^T
\tilde{\gbf} s[k] ~~~\mbox{and}~~~ \hat{y}_{d,N}[k] =
\wbf^H_k\Gbf\Rbf\nbf_r +\wbf^H_k\nbf_d,
\end{equation}
respectively, for $k=0,1,\cdots,N-1$.

\section{Filter-and-Forward Relay Design Criteria and Optimization}
\label{sec:FFrelayProblem}

{In this section, we consider three meaningful FF relay design
problems for the relay network employing OFDM transmission
described in Section \ref{sec:systemmodel}. First, we consider 
the FF relay filter design  to minimize the transmit
power of the FF relay subject to an SNR constraint for each OFDM
subcarrier channel when the source power allocation $\{P_{s,k}\}$ is given.  Here, we shall show that the problem can be formulated as a semi-definite relaxation (SDR) problem. That is, the original non-convex FF relay design
problem is  approximated by a convex SDP problem. Furthermore, in this case  we shall show that the solution to the relaxed SDP
problem is the same as that to the original problem under a 
mild condition. With the formulae for the relay transmit power and the subcarrier channel SNR in terms of the relay filter coefficients and source power allocation obtained to solve the first problem, we next consider two more relay design problems. One is the problem of maximizing the
worst subcarrier SNR subject to source and  relay power constraints and the other is the problem of rate maximization subject to source and  relay power constraints. The first criterion aims not only at  overall quality-of-service (QoS) improvement for subcarrier channels but also at bit error rate (BER) minimization in case of weak or no channel coding.   In the single-user case, bits for one user are distributed across subcarriers and the overall system BER is dominated by the BER of the worst subcarrier channel since the worst error rate dominates the system error rate \cite{Dembo&Zeitouni:book}. Thus, maximization of the worst subcarrier SNR is almost equivalent to minimization of the system BER in the single-user case. For the latter two problems, we  consider joint optimization of the relay filter and  source power allocation. These two problems are non-convex optimization problems with respect to the relay filter tap coefficients and the source power allocation. Thus, it is not easy to find a globally optimal solution. To circumvent this difficulty, we apply alternating optimzation and the projected gradient method to the worst subcarrier SNR maximization problem and the rate maximization problem, respectively, and propose an efficient iterative algorithm for each problem that converges to a locally optimal point at least.}

\subsection{FF relay transmit power minimization under per-subcarrier SNR constraints}

We first consider the problem of designing the FF relay tap
coefficient vector $\rbf=[r_0,r_1,\cdots,r_{L_r-1}]^T$ to minimize
the relay transmit power subject to an SNR constraint per OFDM
subcarrier. This problem is formulated as follows:

\vspace{0.5em}
{\it Problem 1:} For given
source power allocation $\{P_{s,k}, k=0,1,\cdots, N-1\}$,  SR
channel $\fbf$,  RD channel static information $(\sigma_g^2,
L_g)$, FF relay filter order $L_r$, and a set $\{\gamma_k, ~k \in
\Ic\}$ of desired minimum  SNR values for subcarrier channels
$\Ic$, 
 \begin{equation} \label{Problem:RelayPowerMinimization}
 \min_{\rbf}~~ P_r   ~~~~~~~~\mbox{subject to (s.t.)}~~~~ \mbox{SNR}_k \ge \gamma_k, ~~~\forall~~ k \in \Ic \subset \{0,1,\cdots,N-1\}
 \end{equation}
where $P_r$ is the relay transmit power and $\mbox{SNR}_k$ is the
SNR of the $k$-th subcarrier channel.

\vspace{0.5em}

 To solve Problem 1,
we need to express each term in the problem as a function of the
design variable $\rbf$. First, let us derive the SNR on the
$k$-th subcarrier channel in the received signal
(\ref{eq:ReceivedSig_perCarrier}) at the destination. Note that
the signal and noise parts in (\ref{eq:ReceivedSignal}) are
represented in terms of the relay filtering matrix $\Rbf$. The
representation of SNR in terms of $\Rbf$ is redundant since the true variable $\rbf$
is embedded in $\Rbf$. Thus, we need reparameterization of SNR in
terms of $\rbf$, and this can be done based on
(\ref{eq:ReceivedSignal}) by exploiting the Toeplitz structure of
$\Rbf$ as follows. Using (\ref{eq:ReceivedSignal}), we first
express the received signal power at the destination in terms of
$\rbf$ as
 \begin{eqnarray} \label{eq:SignalPower}
\Ebb \{ |\hat{y}_{d,S}[k] |^2  \} &=& N \Ebb \{ \wbf_k^H \Tbf^T
\Fbf^T \Rbf^T \tilde{\gbf} |s[k]|^2 \tilde{\gbf}^H
\Rbf^* \Fbf^* \Tbf^* \wbf_k\},  \nonumber\\
&=& N \wbf_k^H \Tbf^T \Fbf^T \Rbf^T \Ebb \{|s[k]|^2 \tilde{\gbf}
\tilde{\gbf}^H\} \Rbf^* \Fbf^* \Tbf^* \wbf_k,  \nonumber\\
&\stackrel{(a)}{=}& N P_{s,k} \mbox{tr} (\wbf_k^H\Tbf^T \Fbf^T
\Rbf^T \Ebb  \{\tilde{\gbf} \tilde{\gbf}^H\} \Rbf^* \Fbf^* \Tbf^* \wbf_k), \nonumber\\
&\stackrel{(b)}{=}& N P_{s,k} \sigma_g^2 \mbox{tr}
(\wbf_k^H\Tbf^T \Fbf^T \Rbf^T
{\tilde{\Ibf}_{L_g}}  \Rbf^* \Fbf^* \Tbf^* \wbf_k), \nonumber\\%
&\stackrel{(c)}{=}&N P_{s,k} \sigma_g^2 \mbox{tr} (
\underbrace{\Fbf^*\Tbf^*\wbf_k\wbf_k^H \Tbf^T \Fbf^T}_{=:\Kbf_k}
\Rbf_{L_g}^T \Rbf_{L_g}^*), \nonumber\\
&\stackrel{(d)}{=}& N P_{s,k} \sigma_g^2 \mbox{tr}
(\Rbf_{L_g}^*\Kbf_k
\Rbf_{L_g}^T ), \nonumber\\
&\stackrel{(e)}{=}& N P_{s,k} \sigma_g^2
\left[\mbox{vec}(\Rbf_{L_g}^T)\right]^H
\bar{\Kbf}_k \mbox{vec}(\Rbf_{L_g}^T) \nonumber\\
&=& N P_{s,k} \sigma_g^2 \left[\mbox{vec}(\Rbf^T)\right]^H
~\tilde{\Kbf}_k~
\mbox{vec}(\Rbf^T) \nonumber\\
&\stackrel{(f)}{=}& N P_{s,k} \sigma_g^2 \rbf^H \Ebf_1
\tilde{\Kbf}_k \Ebf_1^H \rbf, \label{eq:ydsk_sig_final}
\end{eqnarray}
where
\begin{equation*}
\tilde{\Ibf}_{L_g } :=  \left [\begin{array}{ll}
 \Ibf_{L_g}& {\bf{0}}_{L_g\times (N-1)} \\
 {\bf{0}}_{(N-1) \times L_g} &  {\bf{0}}_{(N-1) \times (N-1)}
\end{array}
 \right ]; ~~~\Rbf = \left [
\begin{array}{l}
\Rbf_{L_g} \\
\Rbf_{N-1}
\end{array}
\right ]; ~~~\bar{\Kbf}_k = \Ibf_{L_g} \otimes \Kbf_k;
~~~\tilde{\Kbf}_k =  \tilde{\Ibf}_{L_g } \otimes \Kbf_k;
\end{equation*}
 $\Rbf_{L_g}$ is a matrix composed of the first $L_g$ rows of $\Rbf$; and
\begin{equation} \label{eq:E1}
\Ebf_1 =  [\underbrace{{\Ibf}_{L_r}, {\bf{0}}_{L_r \times
(N+L_g-2)}}_{N+L_g+L_r-2 ~~\mbox{columns}},
\underbrace{{\bf{0}}_{L_r \times 1} ,{\Ibf}_{L_r} , {\bf{0}}_{L_r
\times (N+L_g-3)}}_{N+L_g+L_r-2 ~~\mbox{columns}}, \cdots,
\underbrace{{\bf{0}}_{L_r \times
(N+L_g-2)},{\Ibf}_{L_r}}_{N+L_g+L_r-2 ~~\mbox{columns}}].
\end{equation}
Here, (a) holds due to the assumption of independence of the signal
and the RD channel coefficients; (b) holds due to the
assumption\footnote{The i.i.d. assumption for $g_l$ is not
necessary. More general cases such as correlated $g_l$ and deterministic $g_l$ can be included in the proposed framework with slight change in the derivation.} of $g_l \stackrel{i.i.d.}{\sim}\Cc\Nc(0,\sigma_g^2)$
(see \eqref{eq:gbftilde}) ; (c) and (d) hold due to $\mbox{tr}(\Abf
\Bbf \Cbf) = \mbox{tr}(\Cbf \Abf \Bbf)$; (e) holds due to $\mbox{tr}
(\Rbf_{L_g}^*\Kbf_k \Rbf_{L_g}^T
)=\left[\mbox{vec}(\Rbf_{L_g}^T)\right]^H \bar{\Kbf}_k
\mbox{vec}(\Rbf_{L_g}^T)$; and (f) is obtained because $\Rbf =
{\mathtt{Toeplitz}}(\rbf^T,$ $N+L_g-1)$ and thus
$\mbox{vec}(\Rbf^T)=\Ebf_1^H \rbf$.   The key point of the
derivation of \eqref{eq:ydsk_sig_final} is that the received
signal power at the $k$-th subcarrier channel is represented as a
quadratic form of the design variable $\rbf$. Next, consider the
received noise power for the $k$-th subcarrier channel. Using
similar techniques to those used to obtain
(\ref{eq:ydsk_sig_final}), we can express the received noise power
based on the noise part in (\ref{eq:ReceivedSignal}) as
\begin{eqnarray}
\Ebb\{|\hat{y}_{d,N}[k] |^2 \} &=& \sigma^2_r {\mbox{tr}}
(\Rbf^H\underbrace{\Ebb\{\Gbf^H \wbf_k\wbf_k^H\Gbf\}}_{=:\Mbf_k} \Rbf ) + \sigma^2_d, \nonumber \\
&=& \sigma^2_r \left[\mbox{vec}(\Rbf)\right]^H \tilde{\Mbf}_k \mbox{vec}(\Rbf)+ \sigma^2_d, \nonumber \\
& = & \sigma^2_r \rbf^H \Ebf_2 \tilde{\Mbf}_k \Ebf_2^H \rbf +
\sigma^2_d, \label{eq:ydnk_noise_power}
\end{eqnarray}
where $\tilde{\Mbf}_k= \Ibf_{N+L_g+L_r-2} \otimes \Mbf_k$ and
$\Ebf_2$ is given by
\begin{equation} \label{eq:Ebf_2form}
\Ebf_2
 = \left [
 \begin{array}{cccccccc}
 \ebf_1^T       &\ebf_2^T      &\cdots &\ebf_{N+L_g-1}^T &{\bf{0}}^T   &\cdots       &\cdots &{\bf{0}}^T\\
 {\bf{0}}^T &\ebf_1^T      &\ebf_2^T &\cdots         &\ebf_{N+L_g-1}^T &{\bf{0}}^T &\cdots &{\bf{0}}^T\\
 {\bf{0}}^T &{\bf{0}}^T&\ddots &\ddots         &\ddots         &\ddots       &\ddots &\vdots  \\
 \vdots       &            &       &\ddots         &\ddots         &\ddots       &\ddots &{\bf{0}}^T   \\
 {\bf{0}}^T &\cdots      &\cdots &{\bf{0}}^T   &\ebf_1^T         &\ebf_2^T       &\cdots & \ebf_{N+L_g-1}^T
 \end{array}
 \right ].
\end{equation}
Here, $\ebf_i^T$ be the $i$-th row of $\Ibf_{N+L_g-1}$ and the
size of  each ${\bf{0}}^T$ in \eqref{eq:Ebf_2form} is $1 \times
(N+L_g-1) $. (It is easy to verify that $\mbox{vec}(\Rbf) =
\Ebf_2^H \rbf$ due to the Toeplitz structure of $\Rbf$.) Based on
\eqref{eq:ydsk_sig_final} and \eqref{eq:ydnk_noise_power}, the SNR
of the $k$-th subcarrier channel is  expressed as
\begin{equation} \label{eq:SNR}
{\mbox{SNR}}_k = \frac{N P_{s,k} \sigma_g^2 \rbf^H \Ebf_1
\tilde{\Kbf}_k \Ebf_1^H \rbf}{\sigma^2_r \rbf^H \Ebf_2
\tilde{\Mbf}_k \Ebf_2^H \rbf + \sigma^2_d}.
\end{equation}
Next, consider the relay transmit power. Using
(\ref{eq:RelayTransmittedSignal}), we obtain the relay transmit
power in a similar way as
\begin{eqnarray}
\Ebb \{  \mbox{tr}( \ybf_t \ybf_t^H) \} &=& \mbox{tr} ( \Rbf
\Fbf\underbrace{\Ebb\{
\tilde{\xbf}_s\tilde{\xbf}_s^H\}}_{=:\Sigmabf_{\tilde{\xbf}_s}}
\Fbf^H\Rbf^H) + \mbox{tr}
(\sigma_r^2 \Rbf \Rbf^H ), \nonumber\\
& = &  \mbox{tr} ( \Rbf
\underbrace{(\Fbf\Sigmabf_{\tilde{\xbf}_s}\Fbf^H +
\sigma_r^2 \Ibf)}_{=:\Pibf} \Rbf^H  ), \nonumber \\
&=& \left[\mbox{vec}(\Rbf^H)\right]^H \tilde{\Pibf} \mbox{vec}(\Rbf^H), \nonumber \\
&=& \rbf^T \Ebf_1 \tilde{\Pibf} \Ebf_1^H \rbf^* = \rbf^H \Ebf_1
\tilde{\Pibf}^* \Ebf_1^H \rbf, \label{eq:RelayPower}
 \end{eqnarray}
where $\tilde{\Pibf} = \Ibf_{N+L_g-1} \otimes \Pibf$, and
$\Sigmabf_{\tilde{\xbf}_s}$ is obtained similarly to
(\ref{eq:covarianceMat}) based on (\ref{eq:CovTildeX}). The last
equality holds since the power is a real-valued quantity.

Now, based on (\ref{eq:ydsk_sig_final}),
(\ref{eq:ydnk_noise_power}) and (\ref{eq:RelayPower}), Problem
1 can be restated  as follows:
\begin{equation} \label{eq:problem1prime}
\underset{\rbf}{\min}   ~~\rbf^H \Ebf_1 \tilde{\Pibf}^* \Ebf_1^H
\rbf  ~~~~~~ \mbox{s.t.} ~~~ \frac{N P_{s,k} \sigma_g^2 \rbf^H
\Ebf_1 \tilde{\Kbf}_k \Ebf_1^H \rbf}{\sigma^2_r \rbf^H \Ebf_2
\tilde{\Mbf}_k \Ebf_2^H \rbf + \sigma^2_d}
 \ge \gamma_k ~~, ~~~~k  \in \Ic.
\end{equation}
The above problem is
not a convex problem. However,
 the problem can still be  solved efficiently by using convex
optimization techniques. Let $\Rc := \rbf\rbf^H$. Then, by using
$\tr(\Abf\Bbf\Cbf)=\tr(\Bbf\Cbf\Abf)$ and relaxing the rank one
constraint for $\Rc$, the problem (\ref{eq:problem1prime}) can
be reformulated as follows:

\vspace{0.5em}
{\it Problem 1$^{\prime}$:}  
\begin{eqnarray} 
&\underset{\Rc}{\min}&  \tr \left ( \Phibf_P \Rc \right )  \label{eq:RelayPowerSDP} \\
&\mbox{s.t.} & \tr \left( [\Phibf_S (k) -\gamma_k \Phibf_N(k)]\Rc \right) \ge \sigma^2_d\gamma_k, ~~~~~ k\in \Ic, \nonumber \\
   &  & \Rc \succeq 0,  \nonumber
\end{eqnarray}
where $\Phibf_P = \Ebf_1 \tilde{\Pibf}^* \Ebf_1^H$, $\Phibf_S (k) =
N P_{s,k} \sigma_g^2 \Ebf_1\tilde{\Kbf}_k \Ebf_1^H$, and $\Phibf_N(k) =
\sigma^2_r\Ebf_2 \tilde{\Mbf}_k\Ebf_2^H$.
\vspace{0.5em}

 Note that by relaxing the rank one constraint for $\Rc$,
Problem 1 is converted to Problem
1$^{\prime}$, which is a semi-definite program (SDP)
\cite{Boyd:04cvxbook} and it can be solved efficiently by using the
standard interior point method for convex optimization
\cite{Boyd:04cvxbook}, \cite{Sturm:99OptSeduMi}. With an
additional constraint $\mbox{rank}(\Rc) = 1$, Problem
1$^{\prime}$ is equivalent to the original Problem
1. That is,  if the optimal
solution to Problem 1$^{\prime}$ has rank one, then
it is also the optimal solution to Problem
1. However, there is no guarantee
that an algorithm for solving Problem 1$^{\prime}$
yields the desired rank one solution. In such a case,
randomization techniques \cite{Sidiropoulos:06SP} can be used to
obtain a rank-one solution $\rbf$ from $\Rc$. However, for this
specific problem related to the  transparent FF relay design, we
provide a stronger result, stated in the following theorem.

\vspace{0.5em}
\begin{theorem} \label{theo:OptimalityRelaxedSDP}
If all the desired SNR constraints except one are satisfied with
strict inequality,
 then the nontrivial optimal solution of Problem 1$^{\prime}$, which is a relaxed version of Problem
1,  always has rank one.
\end{theorem}

\textit{Proof} : See the appendix.

\vspace{0.5em}

 Note that the condition in Theorem
\ref{theo:OptimalityRelaxedSDP} is mild and is satisfied in many
cases.  Thus, solving Problem 1$^{\prime}$ directly
yields the solution to the original power minimization problem
under subcarrier SNR constraints in many cases.

\subsection{Worst subcarrier SNR maximization}
\label{subsec:worstsubSNRmax}

{Now, we consider the second FF relay design problem of  maximizing the SNR
of the worst subcarrier channel under  transmit power
constraints. As mentioned already, this problem is closely related to BER minimization in case of weak or no channel coding in addition to minimum QoS improvement. To be complete, for this important problem we consider not only  relay filter optimization but also optimal source power allocation. The joint optimization yields a further gain over the relay filter optimization only, as seen in other joint optimization \cite{Ng:07JSAC, Stoica:02SP, Palomar:03SP}.  The  problem of joint source power allocation and FF relay
filter design to maximize the worst subcarrier SNR subject to
total source and relay transmit power constraints is formulated as follows:}

\vspace{0.5em}
{\it Problem 2:}
For given SR channel $\fbf$, RD channel statistic information
$(\sigma_g^2, L_g)$, FF relay filter order $L_r$, maximum
available source transmit power $P_{s,max}$, and maximum available
relay transmit power $P_{r,max}$, optimize the relay filter $\rbf$ and the source power allocation $\{P_{s,0}, \cdots, P_{s,N-1}\}$ in order to maximize the worst subcarrier SNR:
\begin{equation} \label{Problem:SNRmax1}
\underset{\rbf, P_{s,0}, \cdots, P_{s,N-1}}{\max} ~~\underset{k \in
\{0, \cdots, N-1\}}{\min} {\mbox{SNR}}_k  ~~~~{\mbox{s.t.}}
~~~~\sum_{k=0}^{N-1}P_{s,k} \le P_{s,max} ~~~\mbox{and}~~~ P_r
\le P_{r,max}.
\end{equation}

\vspace{0.5em}  Note that Problem
2 is a complicated non-convex optimization
problem. There exist several methods that can find the optimal solution of a non-convex optimization problem as long as the cost function is not too complicated  \cite{Horst&Pardalos:book,Pardalos&Romeijin:book}. However, such methods require high computational complexity.  Hence, we here approach the problem by using a suboptimal
alternating optimization technique for computational efficiency. That is, first the  source power allocation
is initialized properly and Problem 2 is solved
to optimize the relay filter for given source power allocation. (This problem is defined as Problem 2-1.)
Then, with the given relay filter tap coefficients obtained by solving
Problem 2-1,  the source power allocation is
optimized. (This problem is defined as Problem 2-2.)  The two problems are solved in an alternating fashion  until the iteration
converges. Let us consider Problem 2-1 first. Problem 2-1 can be written explicitly based on  \eqref{eq:SNR} and  \eqref{eq:RelayPower} as follows:

\vspace{0.5em}

{\it Problem 2-1:}
For given source power allocation $\{P_{s,k}, k=0,1,\cdots,
N-1\}$, SR channel $\fbf$,  RD channel static information
$(\sigma_g^2, L_g)$, FF relay filter order $L_r$, and  maximum
available relay transmit power $P_{r,max}$,
\begin{equation} \label{Problem:SNRmaxOriginal}
\underset{\rbf}{\max}~~ \underset{k \in \{0, \cdots, N-1\}}{\min}
\frac{N P_{s,k} \sigma_g^2 \rbf^H \Ebf_1 \tilde{\Kbf}_k \Ebf_1^H
\rbf}{\sigma^2_r \rbf^H \Ebf_2 \tilde{\Mbf}_k \Ebf_2^H \rbf +
\sigma^2_d} ~~~~~~\mbox{s.t.}~~~
 \rbf^H \Ebf_1 \tilde{\Pibf}^* \Ebf_1^H
\rbf \le P_{r,max} .
\end{equation}

\vspace{0.5em} 
By introducing a slack variable $\tau$, the above max-min problem can be
rewritten as
\begin{eqnarray} 
&\underset{\rbf}{\max} & \tau \label{Problem:SNRmaxOriginalSlack} \\
&{\mbox{s.t.}} & \frac{N P_{s,k} \sigma_g^2 \rbf^H \Ebf_1
\tilde{\Kbf}_k \Ebf_1^H \rbf}{\sigma^2_r \rbf^H \Ebf_2
\tilde{\Mbf}_k \Ebf_2^H \rbf +
\sigma^2_d}\ge  \tau~~, ~~~~~~~~ k = 0, ~1,~ \ldots,~ N-1 \nonumber \\
 & &  \rbf^H \Ebf_1 \tilde{\Pibf}^* \Ebf_1^H
\rbf \le P_{r, max} . \nonumber
\end{eqnarray}
Note that this is a non-convex problem. Again, as in the previous
subsection, we convert the problem to a tractable convex problem by
semi-definite relaxation as follows:
\begin{eqnarray}
&\underset{\Rc}{\max} & \tau \label{eq:ProblemOld5} \\
&{\mbox{s.t.}} & {\mbox{tr}} \left( \left(\Phibf_S (k) -\tau
\Phibf_N(k)\right)
\Rc \right) \ge  \sigma^2_d \tau, ~~~~ k = 0, ~1,~ \ldots,~ N-1 \nonumber \\
& & \mbox{tr} \left ( \Phibf_P \Rc \right ) \le P_{r,max} \nonumber \\
& & \Rc \succeq 0, \nonumber
\end{eqnarray}
where $\Phibf_S(k)$, $\Phibf_N(k)$, and $\Phibf_P$ are already
defined in Problem 1$^{\prime}$.
In the problem (\ref{eq:ProblemOld5}), the rank
constraint $ {\mbox{rank}}(\Rc) = 1$  is dropped by semi-definite
relaxation as in Problem 1$^{\prime}$. Note that the
relaxed optimization problem is quasi-convex, i.e., for given $\tau$
the problem is convex.  The solution of the quasi-convex
optimization problem  can be obtained  by solving its corresponding
feasibility problem \cite{Havary:08SP}:
\begin{eqnarray} 
&\mbox{Find}& \Rc \label{Problem:SNRmaxFeasibility} \\
&{\mbox{s.t.}} & {\mbox{tr}} \left( \left(\Phibf_S (k) -\tau
\Phibf_N(k)\right)
\Rc \right) \ge  \sigma^2_d \tau, ~~~~ k = 0, ~1,~ \ldots,~ N-1 \nonumber \\
& & \mbox{tr} \left ( \Phibf_P \Rc \right ) \le P_{r,max} \nonumber \\
& & \Rc \succeq 0. \nonumber
\end{eqnarray}
The feasible set in the problem (\ref{eq:ProblemOld5})  is convex for
any value of $\tau$. Let $\tau^\star$ be the optimal value of
the problem (\ref{eq:ProblemOld5}). Then, we can find the solution to
the problem (\ref{eq:ProblemOld5})  by using the fact that the
feasibility problem \eqref{Problem:SNRmaxFeasibility} is feasible
for $\tau \le \tau^\star$, whereas it is not feasible for $\tau
> \tau^\star$. Based on this, we propose  a simple bisection algorithm
to solve the problem (\ref{eq:ProblemOld5}), a relaxed version of Problem 2-1, as follows:

\begin{algorithm} \label{algo:bisection} Choose some appropriate interval s.t.  $\tau^\star \in(
\tau_L , \tau_R)$.

\noindent  Step 1: Set $\tau = {(\tau_L + \tau_R)}/{2}$.

\noindent  Step 2:  Solve the feasibility problem
\eqref{Problem:SNRmaxFeasibility}  for $\tau$.  If it is feasible,
$\tau_L = \tau$. Otherwise, $\tau_R = \tau$.

\noindent Step 3: Repeat Steps 1 to 2 until $(\tau_R - \tau_L) <
\epsilon$.
\end{algorithm}

\noindent Here, $\epsilon$ is the allowed error tolerance for
$\tau$. Note that the above feasibility problem is a standard SDP
problem, which can be solved easily by  the interior point method
\cite{Sturm:99OptSeduMi}.  Due to the relaxation, the matrix $\Rc$
obtained by solving the relaxed optimization problem may not have
rank one in general.  In such a case, randomization techniques can
be applied to find a rank-one solution.

When we optimize only the relay filter  to maximize the worst subcarrier SNR for given source power allocation, we can simply use Algorithm \ref{algo:bisection} only. However, for joint optimization of the relay filter and source power allocation by alternating optimization, we need to consider Problem 2-2, which is given as follows:

\vspace{0.5em}

{\it Problem 2-2:}
For given FF relay filter $\rbf$, SR channel $\fbf$,  RD channel
statistic information $(\sigma_g^2, L_g)$, maximum allowed source
transmit power $P_{s,max}$, and maximum allowed relay transmit
power $P_{r,max}$,
\begin{equation}  \label{Problem:JointProblem2}
\underset{P_{s,0}, \cdots, P_{s,N-1}}{\max} ~~\underset{k \in \{0,
\cdots, N-1\}}{\min} {\mbox{SNR}}_k  ~~~{\mbox{s.t.}}
~~\sum_{k=0}^{N-1}P_{s,k} \le P_{s,max}, ~~P_r \le P_{r,max},
~~\mbox{and}
 ~~{\mbox{SNR}}_k \ge \tau_0 ~~\forall~k,
 \end{equation}
where {$\tau_0$ is the allowed minimum for the worst subcarrier SNR}.

\vspace{0.5em} 

 The constraint ${\mbox{SNR}}_k \ge \tau_0 ~~\forall~k$ in (\ref{Problem:JointProblem2}) is introduced intentionally to guarantee that the proposed alternating algorithm yields a monotone non-decreasing sequence of the worst subcarrier SNR values. (This will become clear shortly.)  By introducing a slack variable $\tau$
and using \eqref{eq:SNR} and  \eqref{eq:RelayPower}, Problem
2-2  can be rewritten   as follows:
\begin{eqnarray}
&\underset{P_{s,0}, \cdots, P_{s,N-1}, \tau}{\max} & \tau  \label{eq:jointoptSP} \\
 &{\mbox{s.t.}} & \sum_{k=0}^{N-1}P_{s,k} \le P_{s,max}, \nonumber \\
 & &  \rbf^H \Ebf_1 \tilde{\Pibf}^* \Ebf_1^H
\rbf \le P_{r,max}, \label{constraint:relaypower} \\
& & \frac{N P_{s,k} \sigma_g^2 \rbf^H \Ebf_1 \tilde{\Kbf}_k \Ebf_1^H
\rbf}{\sigma^2_r \rbf^H \Ebf_2 \tilde{\Mbf}_k \Ebf_2^H \rbf +
\sigma^2_d}\ge  \tau~~, ~~~~~~~~ k = 0, ~1,~ \ldots,~ N-1, \nonumber \\
 & & \tau \ge \tau_0. \nonumber 
\end{eqnarray}
Note that without the relay power constraint
\eqref{constraint:relaypower}, the above problem is a simple linear
programming (LP) with respect to $P_{s,0}, \cdots, P_{s,N-1}$ and $
\tau$. Indeed, the problem is an LP since the relay power constraint
can also be written as a linear form in terms of $P_{s,k}$, as shown
below. The relay power \eqref{eq:RelayPower} can be rewritten as
\begin{eqnarray}
\Ebb \{  \mbox{tr}( \ybf_t \ybf_t^H) \} &=& \mbox{tr} ( \Rbf
\Fbf\Ebb\{ \tilde{\xbf}_s\tilde{\xbf}_s^H\} \Fbf^H\Rbf^H) +
\mbox{tr}
(\sigma_r^2 \Rbf \Rbf^H ), \nonumber\\
& = &  \mbox{tr} ( \Ebb\{ \tilde{\xbf}_s\tilde{\xbf}_s^H\}
\Fbf^H\Rbf^H\Rbf\Fbf) + \mbox{tr}
(\sigma_r^2 \Rbf \Rbf^H ), \nonumber\\
& \stackrel{(a)}{=}&  \mbox{tr} ( \tilde{\Wbf}_N~\Ebb\{\sbf
\sbf^H\} ~\tilde{\Wbf}_N^H \Fbf^H\Rbf^H\Rbf\Fbf) + \mbox{tr}
(\sigma_r^2 \Rbf \Rbf^H ), \nonumber\\
& \stackrel{(b)}{=}&  \sum_{k=0}^{N-1}P_{s,k} ~\mbox{tr} (
~\ebf_{k+1} \ebf_{k+1}^T ~\tilde{\Wbf}_N^H
\Fbf^H\Rbf^H\Rbf\Fbf~\tilde{\Wbf}_N) + \mbox{tr} (\sigma_r^2 \Rbf
\Rbf^H ) \label{constraint:relaypower2}
 \end{eqnarray}
where $\ebf_k$ is defined in \eqref{eq:Ebf_2form}, and
$\tilde{\Wbf}_N$ is the cyclic prefix extended IDFT matrix given by
\begin{equation*}
\tilde{\Wbf}_N = [ \wbf_{N-1} ,~ \wbf_{N-2},~ \cdots, ~\wbf_{0},
~\wbf_{N-1}, \cdots, ~\wbf_{N-L_g-L_r-L_f+3}]^T.
\end{equation*}
Here, (a) can be verified by using \eqref{eq:xs_tilde} and (b) is
due to the assumption of $s[k] \sim {\mathcal{CN}} (0, P_{s,k})$
for $k=0,1,\cdots,N-1$. Using the new expression
\eqref{constraint:relaypower2} for the relay transmit power,  we
obtain an LP optimization problem for the source power allocation
from the problem (\ref{eq:jointoptSP}) as
\begin{eqnarray}
&\underset{P_{s,0}, \cdots, P_{s,N-1} , \tau}{\max} & \tau \label{eq:JointIntProbFinal} \\
 &{\mbox{s.t.}} & \sum_{k=0}^{N-1}P_{s,k} \le P_{s,max},  \nonumber\\
 & &  \sum_{k=0}^{N-1}P_{s,k} ~C_1(k) + C_2 \le P_{r,max} \label{eq:JointIntProbFinalRelPowerC}\\
 & &P_{s,k} ~C_3(k) \ge \tau~~, ~~~~~~~~ k = 0, ~1,~ \ldots,~ N-1,  \nonumber\\
 & & \tau \ge \tau_0, \nonumber
\end{eqnarray}
where $C_1(k) = \mbox{tr} ( ~\ebf_{k+1}
\ebf_{k+1}^T ~\tilde{\Wbf}_N^H
\Fbf^H\Rbf^H\Rbf\Fbf~\tilde{\Wbf}_N)$, $C_2 = \mbox{tr}
(\sigma_r^2 \Rbf \Rbf^H )$, and $C_3(k)=\frac{N \sigma_g^2 \rbf^H \Ebf_1 \tilde{\Kbf}_k
\Ebf_1^H \rbf}{\sigma^2_r \rbf^H \Ebf_2 \tilde{\Mbf}_k \Ebf_2^H
\rbf + \sigma^2_d}$.  Since the problem (\ref{eq:JointIntProbFinal}) is a  LP problem, Problem 2-2 can easily be solved by a standard convex optimization solver.

Now, combining Problems 2-1 and 2-2, we present our alternating
optimization algorithm for the joint source power allocation and relay filter
design problem  to maximize the worst
subcarrier SNR, given in Algorithm \ref{algo:worstSNRjoint}.

\begin{algorithm} \label{algo:worstSNRjoint}  Given parameters:
$\fbf$, $(\sigma_g^2, L_g)$,  $L_r$,  $P_{s,max}$, and
$P_{r,max}$.

\noindent Step 1: Initialize $P_{s,k}$ for $k = 0, \cdots, N-1$. For example, $P_{s,k}=P_{s,max}/N$.

\noindent Step 2: Solve Problem 2-1 with
Algorithm \ref{algo:bisection}.

\noindent Step 3: {Set the allowed minimum $\tau_0$ for the worst subcarrier SNR in Problem 2-2 as the maximum value $\tau^\star$
obtained from Algorithm \ref{algo:bisection} in Step 2.}

\noindent Step 4: For given $\rbf$ and $\tau_0$ from Steps 2 and 3,  solve
Problem 2-2    by solving the problem \eqref{eq:JointIntProbFinal} to obtain new $ P_{s,k}$, ~$k =
0, \cdots, N-1$.

\noindent Step 5:   Go to Step 2. Here,  set $\tau_L$ of Problem 2-1 as  the solution to
Problem \eqref{eq:JointIntProbFinal} in Step 4.

\noindent Step 6: Repeat Steps 2 to 5 until  $|\tau_0 - \tau_L| <
\epsilon$.

\end{algorithm}

\vspace{0.5em}   Here, $\epsilon$ is the allowed error
tolerance for $\tau$. {Note that at each iteration the value $\tau^\star$ of the worst subcarrier SNR is monotone non-decreasing. This is because the maximum value of the previous step is set as a lower bound of the current step and the problem at the current step is feasible since the previous combination of $\{P_{s,k}\}$ and $\rbf$ achieves the current lower bound. Since $\tau^\star$ is monotone non-decreasing and upper bounded because of finite transmit power $P_{s,max}$ and $P_{r,max}$, the proposed algorithm converges to a locally optimal point by the monotone convergence theorem.}  {Although convergence to the global
optimum is not  guaranteed,} it will be seen in Section \ref{sec:NumericalResult} that
the proposed joint design approach improves the  performance
significantly over the relay filter optimization only.

\subsection{{Rate maximization}}
\label{subsec:ratemax}

The third design criterion that we consider in this paper is rate maximization. This problem is especially interesting when high data rates are the main goal of the system design. Again for this rate maximization problem, we consider joint optimization of the relay filter and source power allocation. Based on the expressions for the subcarrier SNR and the relay power obtained in the previous subsections, the problem is formulated as follows:

\vspace{0.5em}
{\it Problem 3:} ~~
For given  $\fbf$,
 $(L_g,\sigma_g^2)$,
 $L_r$,  $P_{s,max}$, and  $P_{r,max}$,
\begin{eqnarray}
&\underset{\rbf, P_{s,0}, \cdots, P_{s,N-1}}{\max} &
\sum_{k=0}^{N-1}  \log \left( 1+   \frac{N P_{s,k}
\sigma_g^2 \rbf^H \Ebf_1 \tilde{\Kbf}_k \Ebf_1^H \rbf}{\sigma^2_r
\rbf^H \Ebf_2 \tilde{\Mbf}_k \Ebf_2^H \rbf +
\sigma^2_d}\right)  \label{eq:jointoptRate} \\
 &{\mbox{s.t.}} & \sum_{k=0}^{N-1}P_{s,k} \le P_{s,max},  \\
 & &  \rbf^H \Ebf_1 \tilde{\Pibf}^* \Ebf_1^H
\rbf \le P_{r,max}. \label{constraint:relaypowerRate}
\end{eqnarray}

\vspace{0.5em}  
When the relay filter is given, an optimal solution to the problem is simply given by  the well-known water-filling strategy for parallel Gaussian channels \cite{Cover&Thomas:book}. However, the freedom to design the relay filter and the dependence of the relay transmit power on the source power allocation make the problem far more difficult than a simple water-filling problem. Note that with $(\rbf,P_{s,0},\cdots,P_{s,N-1})$ as the design variable, the problem is a non-convex problem. Due to the structure of the cost function, it is not easy to convert the problem to a certain convex problem as in the previous subsections. Thus, as in \cite{KimSungLee:12SP}, we adopt a direct numerical method to solve this problem based on the projected gradient method (PGM) which consists of a gradient descent step for cost reduction and a projection onto the constraint set at each iteration and is widely used for constrained optimization \cite{goldstein64, Polyak:69USSR, Slavakis&Yamada&Ogura:06NFAO, KimSungLee:12SP}.   To apply the PGM, we rewrite Problem 3  as follows:  
\begin{eqnarray}
&\underset{\rbf, P_{s,0}, \cdots, P_{s,N-1}}{\min} &
-\sum_{k=0}^{N-1}  \log \left( 1+   \frac{P_{s,k}
\rbf^H \Qbf_1(k) \rbf}{\rbf^H \Qbf_2(k)\rbf + \sigma^2_d}\right)  \label{eq:jointoptRate2} \\
 &{\mbox{s.t.}} & {\bf{1}^T}  \pbf \le P_{s,max},  \label{constraint:sourcepowerRate2}\\
 & &  \rbf^H \Ebf_1 \tilde{\Pibf}^* \Ebf_1^H \rbf \le P_{r,max}. \label{constraint:relaypowerRate2}
\end{eqnarray}
where $\Qbf_1(k) = N \sigma_g^2 \Ebf_1 \tilde{\Kbf}_k \Ebf_1^T $,
$\Qbf_2(k) = \sigma^2_r  \Ebf_2 \tilde{\Mbf}_k \Ebf_2^H $, $\pbf = [P_{s,0},\cdots,P_{s,N-1}]^T$,  and
${\bf{1}} = [ 1 ,\cdots , 1]^T$ .
Then,  the joint design variable vector $\ubf$ and the cost function for the PGM are respectively given by
\begin{eqnarray}
\ubf &:=&[\pbf^T,\rbf^T]^T,\\
\phi(\ubf) &:=& -\sum_{k=0}^{N-1}  \log \left( 1+
\frac{P_{s,k} \rbf^H \Qbf_1(k) \rbf}{\rbf^H \Qbf_2(k)\rbf +
\sigma^2_d}\right).
\end{eqnarray}
The gradient of $\phi(\ubf)$ w.r.t. $\ubf$ can be
obtained as
\begin{equation}\label{eq:gradient}
\phi'(\ubf) = -\frac{1}{ \ln
2}\sum_{k=0}^{N-1}\frac{1}{1+P_{s,k}\frac{{\mathcal{B}}_1(k)}{{\mathcal{B}}_2(k)}}\cdot\frac{1}{{\mathcal{B}}_2(k)^2}
\left[\begin{array}{cc}
\ebf_{k+1}{\mathcal{B}}_1(k) {\mathcal{B}}_2(k) \\
P_{s,k}({\mathcal{B}}_2(k) - {\mathcal{B}}_1(k)){\mathcal{B}}_3(k)
\end{array}
\right ]
\end{equation}
where ${\mathcal{B}}_1(k) = \rbf^H \Qbf_1(k) \rbf$ ,
$~{\mathcal{B}}_2(k) = \rbf^H \Qbf_2(k)\rbf + \sigma^2_d$, and
${\mathcal{B}}_3(k) = (\Qbf_1(k)+\Qbf_1(k)^T) \rbf $. The constraint set $\Sc_1$ defined by   \eqref{constraint:sourcepowerRate2} is  a half-space ${\Hc}_{\pbf}$ for $\pbf$ defined by a hyperplane with no restriction on $\rbf$  and thus $\Sc_1$ is a convex set of $\ubf$.  However, the constraint set $S_2$ defined by \eqref{constraint:relaypowerRate2} is not convex but 
 biconvex w.r.t. $\pbf$ and $\rbf$. That is, $\Sc_2$  is an ellipsoid  $\xi_{\rbf}(\pbf)$ for
$\rbf$ for given $\pbf$  as seen in \eqref{constraint:relaypowerRate2} and is a
half-space for $\pbf$ for given $\rbf$ as seen in (\ref{eq:JointIntProbFinalRelPowerC}). Thus, projection onto  $\Kc := \Sc_1 \cap \Sc_2$ can  be implemented effectively by successive projections: one projecting $\pbf$ onto $\Hc_\pbf$ and the other projecting $\rbf$ onto  the ellipsoid of $\rbf$ for the given projected $\pbf$ by the first projection.   Based on these projections and \eqref{eq:gradient}, we can apply the PGM to Problem 3 in a similar way to that in  \cite{KimSungLee:12SP}. 
  It is guaranteed that the PGM yields a unique globally optimal solution when it is applied to a convex optimization problem \cite{Slavakis&Yamada&Ogura:06NFAO}. However, Problem 3 is not a convex problem, and thus the proposed algorithm does not guarantee convergence to a globally optimal point.  However, numerical results show that the algorithm converges and works well.

\section{Numerical results}
\label{sec:NumericalResult}

 In this section, we provide some numerical results to evaluate the
 performance of the
 FF relay design methods proposed in Section
\ref{sec:FFrelayProblem}. We considered a relay network with an
OFDM transmitter, an FF relay, and a destination node, as described
in Section \ref{sec:systemmodel}. Throughout the simulation, we fixed the number  of OFDM subcarriers
 as $N=32$ with a minimal cyclic prefix covering the overall
FIR channel length in each simulation case.  In all cases, both SR and RD channel tap coefficients $f_l$'s and $g_l$'s were
generated i.i.d. according to the Rayleigh distribution, i.e., $f_l
\stackrel{i.i.d.}{\sim} {\cal{CN}}(0, 1)$ for $l = 0, 1, \cdots,
L_f-1$ and $g_l \stackrel{i.i.d.}{\sim} {\cal{CN}}(0, 1)$ for $l =
0, 1, \cdots, L_g-1$; the relay and the
destination had the same noise power $\sigma_r^2 = \sigma_d^2 = 1$;
and the source transmit power was 20 dB higher than the noise
power, i.e., {$P_{s,max}=100$. (From here on, all  dB power values are relative to $\sigma_r^2=\sigma_d^2=1$.)}

\begin{figure}[ht]
\centerline{
\SetLabels
\L(0.25*-0.1) (a) \\
\L(0.75*-0.1) (b) \\
\endSetLabels
\leavevmode
\strut\AffixLabels{
\scalefig{0.5}\epsfbox{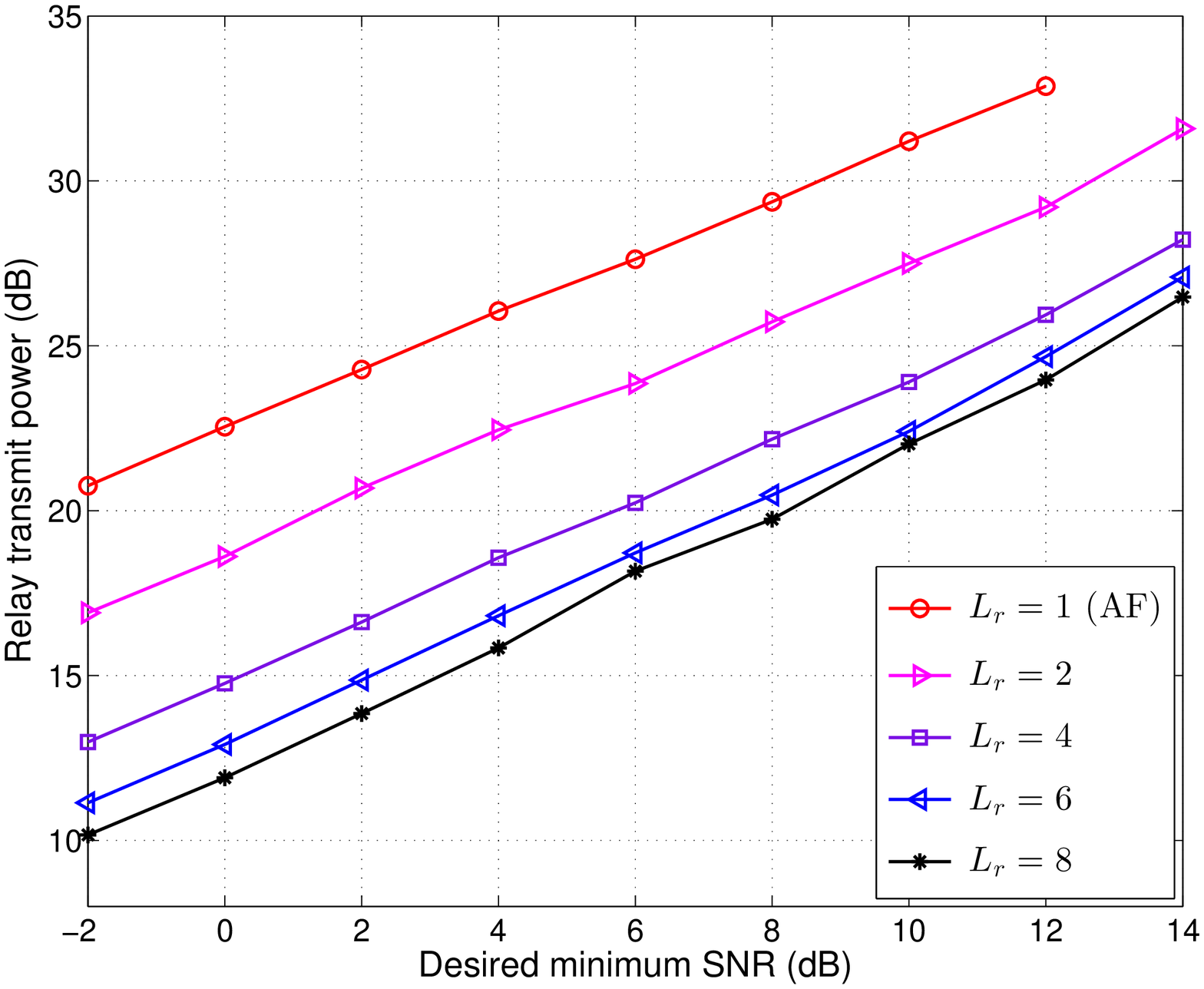}
\scalefig{0.5}\epsfbox{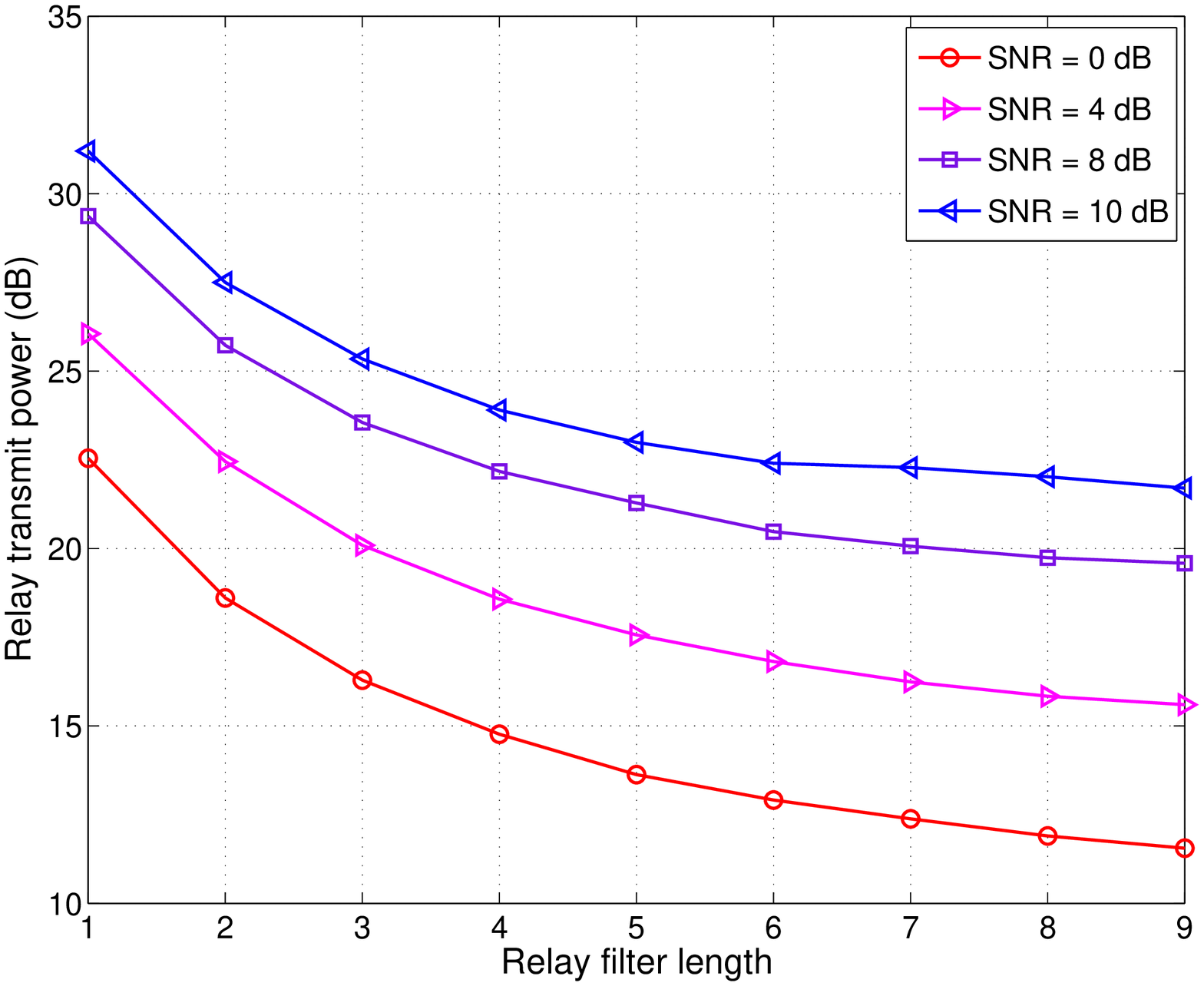} } }
\vspace{1.5em} \caption{FF relay transmit power minimization ({$P_{s,max}=20$ dB, $P_{s,k}=P_{s,max}/N$, $L_f=L_g=3$}): (a) the relay transmit power versus the desired minimum SNR and (b) the relay transmit power versus the FF relay filter length $L_r$ ($L_r=1$:AF)} \label{fig:relayPowerMin}
\end{figure}

We first examined the performance of the first FF relay design method, provided in Problem
1$^\prime$, to  minimize the relay transmit power subject to required SNR constraints on subcarrier channels.
Fig. \ref{fig:relayPowerMin} shows the corresponding result. 
Here, the SR channel length and the RD channel length were set as $L_f = L_g = 3$. 
\vspace{0.1cm}

\begin{figure}[ht]
\centerline{
\SetLabels
\L(0.25*-0.1) (a) \\
\L(0.75*-0.1) (b) \\
\endSetLabels
\leavevmode
\strut\AffixLabels{
\scalefig{0.5}\epsfbox{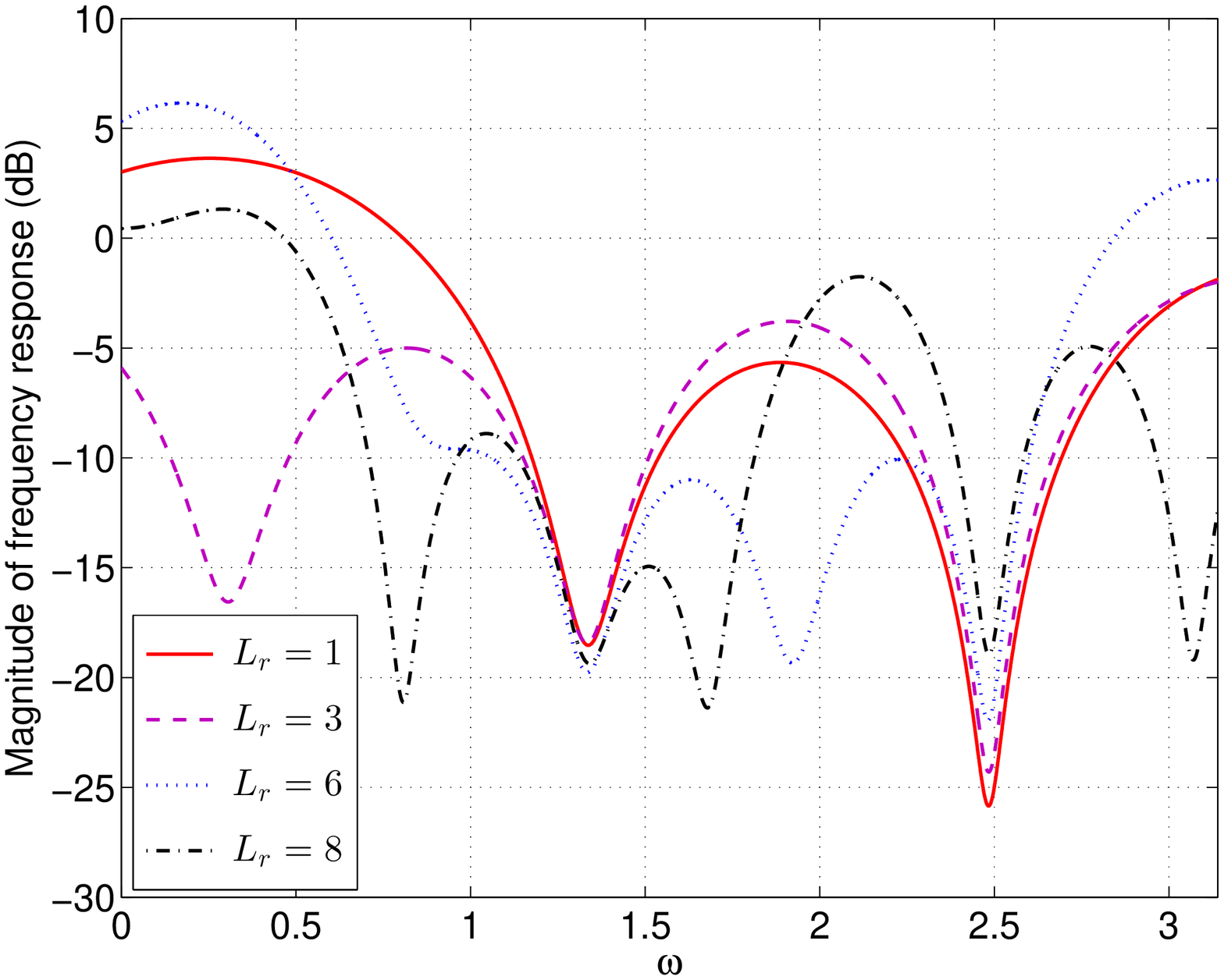}
\scalefig{0.5}\epsfbox{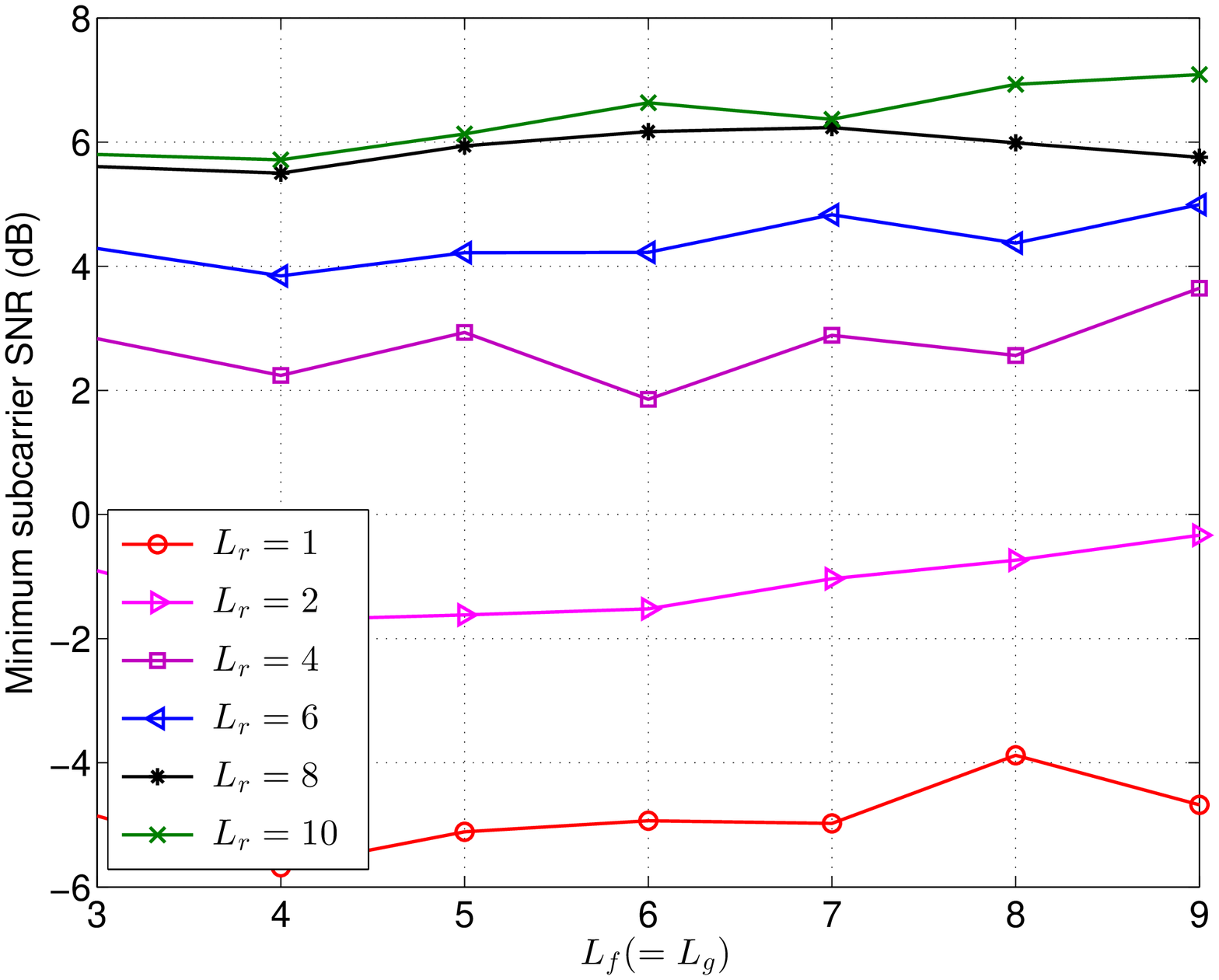} } }
\vspace{1.5em} \caption{{$P_{s,max}=$20 dB, $P_{r,max}=20$ dB,  $P_{s,k}=P_{s,max}/N$, and relay filter optimization only by Algorithm 1: (a) frequency response  and (b) impact of the channel order $L_f=L_g$}} \label{fig:FrequencyRes}
\end{figure}

\noindent We chose $\Ic =\{0,1,\cdots, 27\}$ from 32 subcarriers. It is known
that for a set of randomly realized propagation channels, it is
not easy to always guarantee the desired SNR for every subcarrier
channel when the desired SNR value is high \cite{Chen10:SP}. Thus,
in Figures  \ref{fig:relayPowerMin} (a) and (b), each line was plotted when
Problem 1$^\prime$ was feasible for more than 50
\% out of 1000 random channel realizations for the given minimum
required SNR value for all the subcarrier channels in $\Ic$, and the
plotted value is the relay transmit power averaged over the feasible
channel realizations. It is seen that the required relay transmit
power for the same minimum SNR required by the FF relay is
significantly reduced when compared to that required by the AF relay.  
 Fig.
\ref{fig:relayPowerMin} (b) shows the relay transmit power versus
the relay filter length $L_r$ for various desired minimum SNR
values. It is seen that the required relay transmit power for the
same desired minimum SNR decreases monotonically with respect to
$L_r$, as expected, and the FF relay achieves most of the gain with only a few FF filter taps.
\begin{figure}[http]
\centerline{
\SetLabels
\L(0.25*-0.1) (a) \\
\L(0.75*-0.1) (b) \\
\endSetLabels
\leavevmode
\strut\AffixLabels{
\scalefig{0.5}\epsfbox{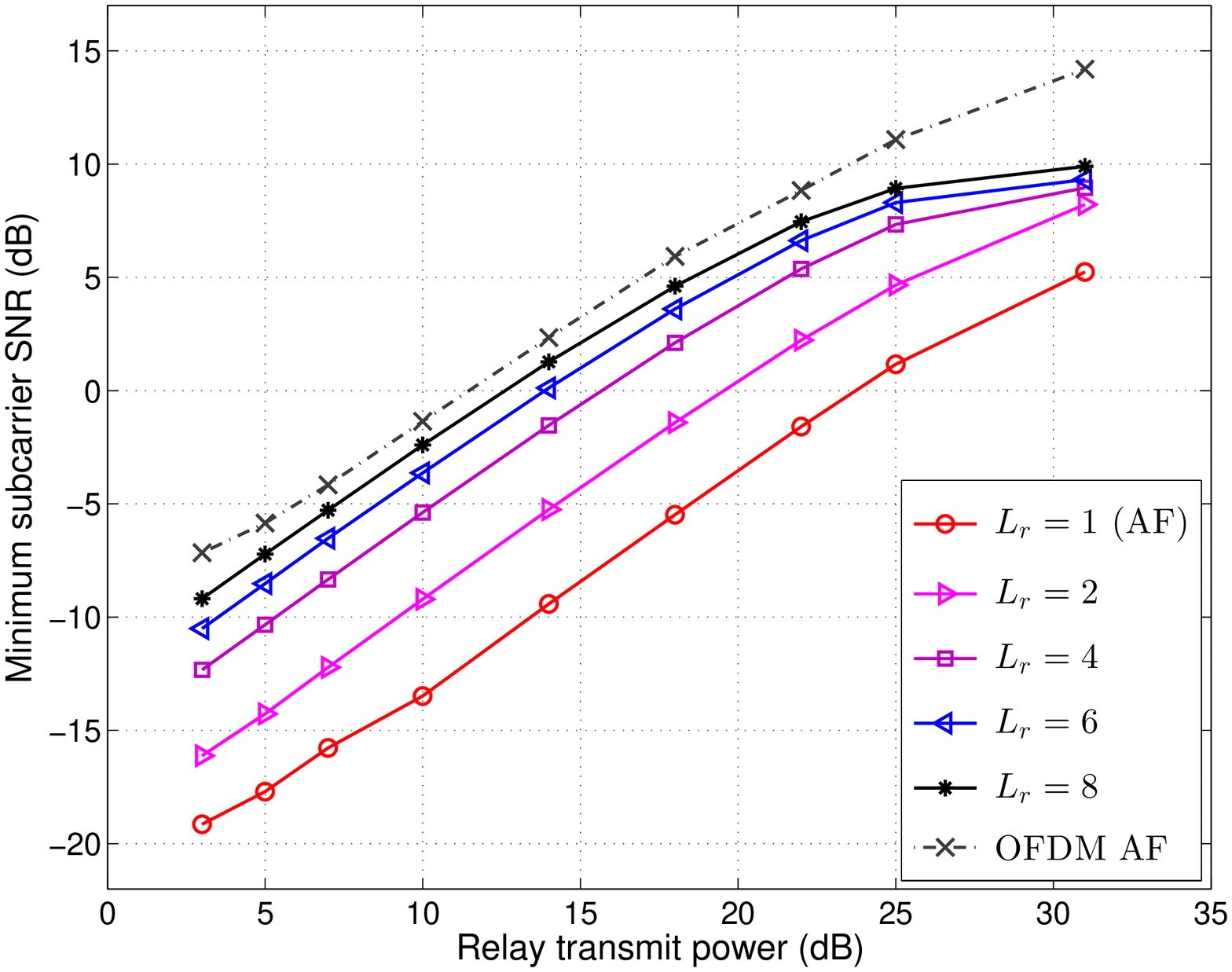}
\scalefig{0.5}\epsfbox{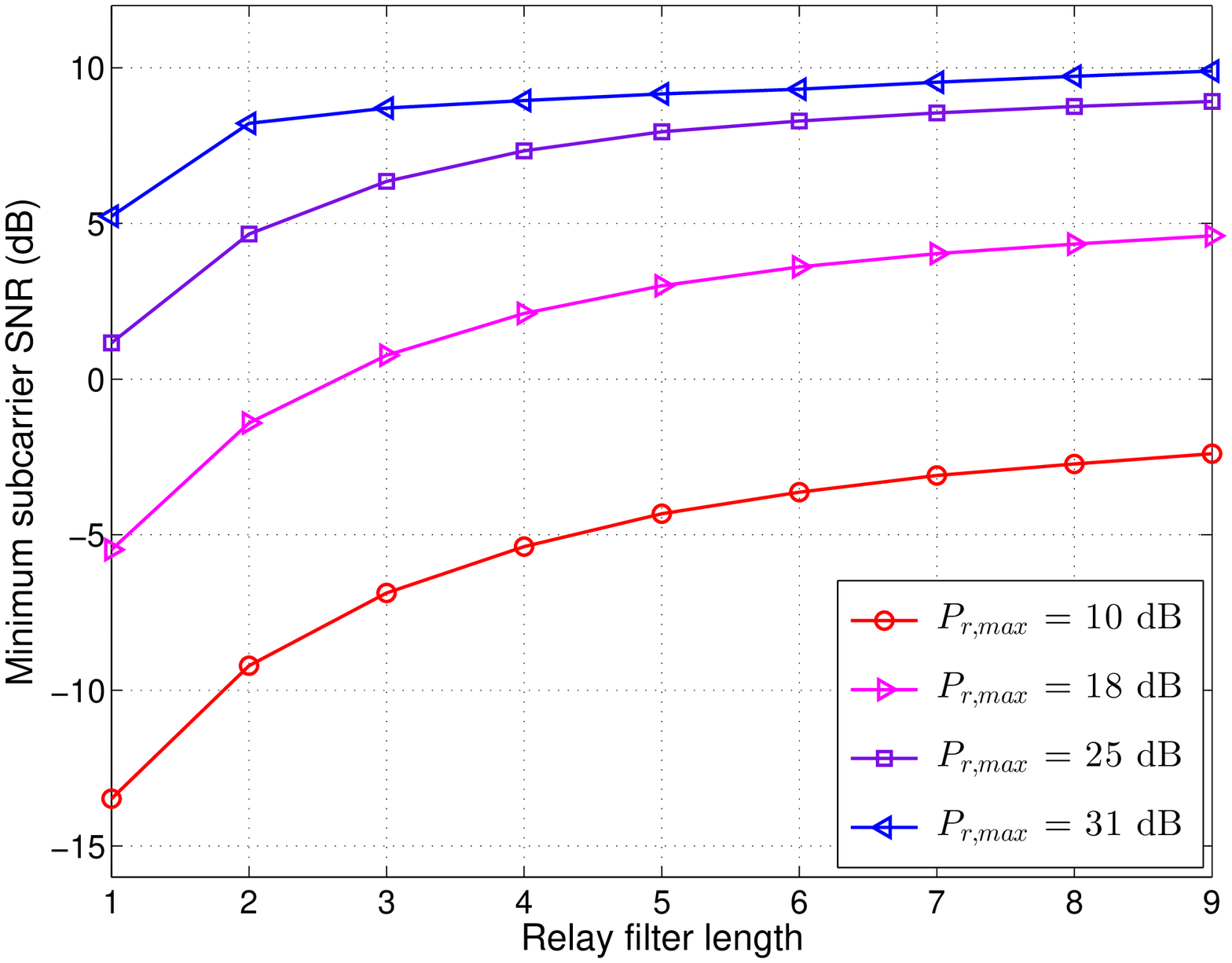} } }
\vspace{1.5em} \caption{The worst subcarrier SNR maximization - Algorithm 1 (relay filter optimization only) ({$P_{s,max}=$20 dB, $P_{s,k}=P_{s,max}/N$,  $L_f=L_g=3$}): (a) the worst subcarrier SNR versus the relay transmit power $P_{r,max}$ and (b) the worst subcarrier SNR versus the FF relay filter length $L_r$ (AF:$L_r=1$)} \label{fig:worstSubCarSNRMax}
\end{figure}
Next, we evaluated the performance of the second FF relay
design method to
maximize the worst subcarrier SNR subject to  transmit
power constraints. First, we considered the relay filter optimization only {for given equal  source power allocation, i.e., $P_{s,k}=P_{s,max}/N$, based on Algorithm 1.} Fig. \ref{fig:worstSubCarSNRMax} shows the result. For the figure, 500 channels were randomly realized with $L_f=L_g=3$ and each plotted value is the average over the 500 channel realizations.  Here, an  OFDM-processing per-subcarrier AF relay is used as an upper bound of the FF relay. {\footnote{The derivation of the OFDM-processing per-subcarrier AF relay design for the worst subcarrier SNR maximization is available at http://wisrl.kaist.ac.kr/papers/wisrltechrep2013feb01.pdf.}} As in the previous case of relay transmit power
minimization, the gain by the FF relay over the AF relay ($L_r=1$) is
significant. Note in Fig. \ref{fig:worstSubCarSNRMax} (a) that the performance of the FF relay improves as the FF relay filter length increases, and eventually converges to the performance of the OFDM-processing per-subcarrier AF relay designed for the same objective in the range of low and intermediate relay power. This is because what the FF relay does is  spectral shaping of the overall channel  essentially (see Fig. \ref{fig:FrequencyRes} (a)) and this spectral shaping can be done maximally with the OFDM-processing per-subcarrier AF relay. Note that most of the gain is achieved by only a few filter
taps for the FF relay and the performance of the FF relay approaches the upper bound quickly in the range of low and intermediate relay power.  On the other hand,  it is seen  in Fig. \ref{fig:worstSubCarSNRMax} (a) that the performance of the FF relay saturates in the high relay transmit power range for $L_r=4,6,8$. At high SNR, precise filtering is required to put the relay power exactly on channel notches to maximize the worst-subcarrier SNR and the situation is much more strict than the low SNR case in which the channel notches are immersed in the noise floor. The observed increased gap between the OFDM-processing relay and the FF relay in the high relay power range implies that the proposed algorithm can  be stuck at some local optimum not at the exact filtering point easily in the the high relay power range. However, the FF scheme still provides far better perfomance than the AF scheme even in this case, and furthermore the practical operating SNR may not be so high to experience such a saturation problem.\footnote{Recall that with BPSK or QPSK the required SNR values for uncoded BER 10$^{-3}$ and 10$^{-6}$ are 6.8 dB and 10.5 dB, respectively. Thus, the saturation around 10 dB minimum SNR may not cause a problem in practical situations.}  
\begin{figure}[http]
\centerline{
\SetLabels
\L(0.25*-0.1) (a) \\
\L(0.75*-0.1) (b) \\
\endSetLabels
\leavevmode
\strut\AffixLabels{
\scalefig{0.5}\epsfbox{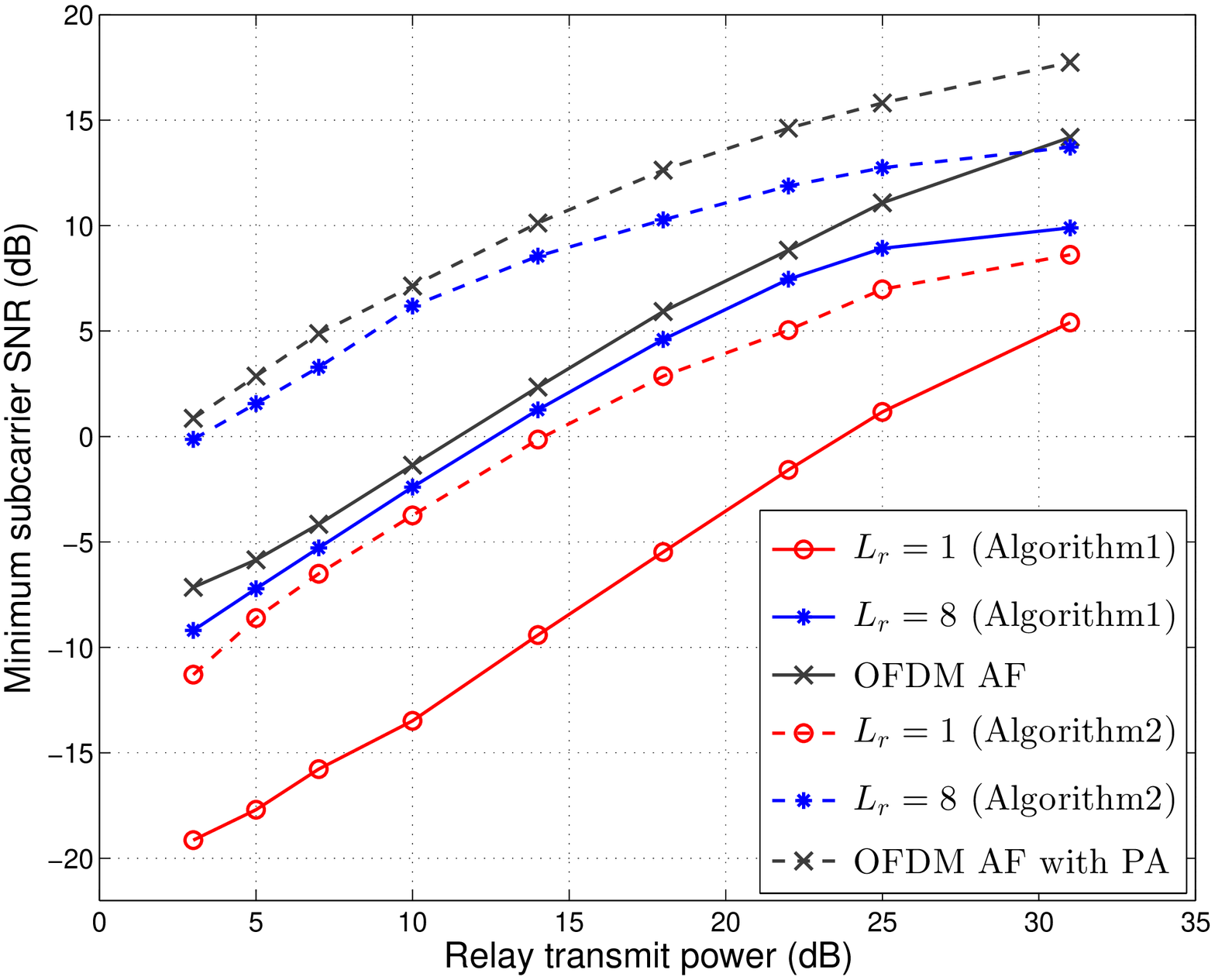}
\scalefig{0.5}\epsfbox{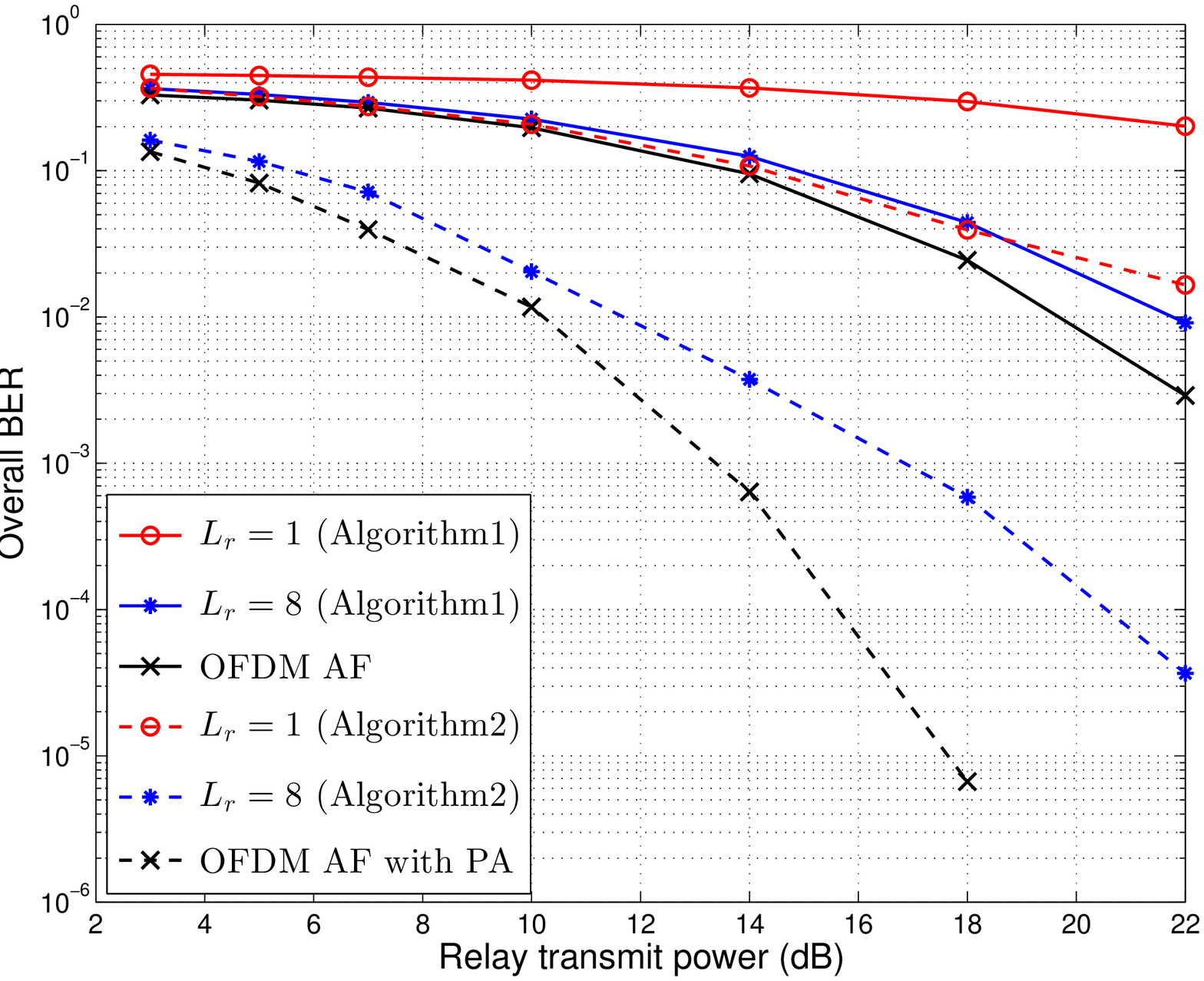} } }
\vspace{1.5em} \caption{$P_{s,max}=$20 dB,  $L_f=L_g=3$: (a) the worst subcarrier SNR versus the relay transmit power $P_{r,max}$  and (b) Overall BER versus $P_{r,max}$} \label{fig:worstSubCarSNRMaxJoint}
\end{figure}
Fig. \ref{fig:FrequencyRes} (a) shows several frequency responses of interest for a set of randomly realized channel vectors $\fbf$ and $\gbf$ with $L_f=L_g=3$. ($\fbf=[ -0.0477 + 0.7546i, ~~   0.1938 + 0.2019i, ~~  -0.4832 - 0.2111i]$ and $\gbf=[-0.8370 - 0.2463i, ~~  -0.3438 + 0.1734i,~~  -0.5136 + 0.4147i]$.) The frequency response of $f[l]*g[l]$ is the frequency response of the original channel from the source to the destination and the frequence response of $f[l]*r[l]*g[l]$ is the shaped channel response by the relay filter $r[l]$ desinged by Algorithm 1. Note that the notches of the original channel response are filled by the frequency shaping by the relay filter in order to maximize the worst subcarrier SNR and this shaping elaborates as the filter order increases. Fig. \ref{fig:FrequencyRes} (b) shows the impact of the channel order $L_f=L_g$. From Fig. \ref{fig:FrequencyRes} (b), it is deduced that the overall channel order of 5 with $L_f=L_g=3$ already presents quite complicated frequency selectivity and therefore the complexity of frequency selectivity caused by higher channel orders does not impact much on the worst subcarrier SNR maximization when the minimum required SNR is not too high.

  We next evaluated the performance of the joint source power allocation
and  FF relay filter design method to maximize the worst subcarrier
SNR, provided in Algorithm 2, and the  result is shown in Fig. \ref{fig:worstSubCarSNRMaxJoint}.   Again, the  OFDM-processing per-subcarrier AF relay was used as a performance upper bound. It is seen  that
the joint optimization method significantly outperforms the
optimization of the relay filter only presented in Algorithm 1. As mentioned in the previous section, the worst subcarrier SNR maximization is closely related to BER minimization. We investigate the BER performance corresponding to Fig. \ref{fig:worstSubCarSNRMaxJoint} (a) and the result is shown in Fig. \ref{fig:worstSubCarSNRMaxJoint} (b).
{Here,  we assumed uncoded QPSK modulation for each subcarrier channel. From the result of Fig. \ref{fig:worstSubCarSNRMaxJoint} (a), we knew the SNR of each subcarrier channel of the total $N=32$ subcarrier channels for the designed FF relay filter and source power allocation. Based on this, we  computed the subcarrier BER based on the  SNR of each subcarrier and averaged all the subcarrier channel BERs to obtain the overall BER. Although there is a noticeable degradation  from the OFDM-processing per-subcarrier AF relay at the high SNR, the FF relay significantly improves the BER performance over the  AF relay.} 

\begin{figure}[http] \centerline{
    \begin{psfrags}
      \scalefig{0.5}\epsfbox{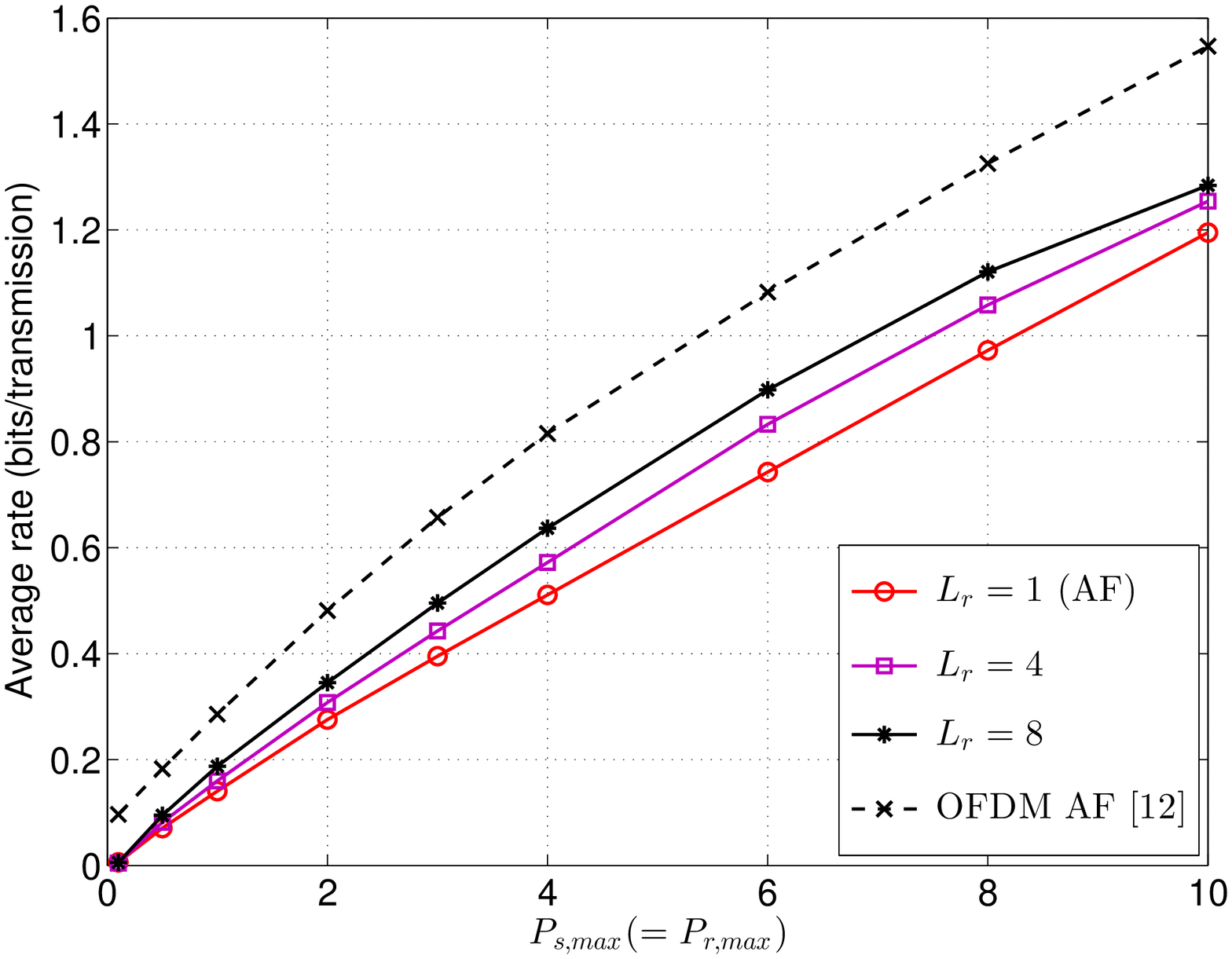}
    \end{psfrags}
} \caption{Sum rate (averaged over 200 channel realizations) versus $P_{s,max}=P_{r,max}$: $L_f=L_g=3$ }
\label{fig:RateMax}
\end{figure}

Next, we evaluated the performance of the third FF relay
design method, provided in Section III.C,  to
maximize the sum rate of subcarrier channels subject to  source and relay transmit
power constraints. The result is shown in Fig. \ref{fig:RateMax}. As a performance upper bound, we considered the method in \cite{Hammerstr:06ConfCommun} which does full OFDM processing and subcarrier reordering for rate maximization with full knowledge of the SR and RD channel state. It is observed that the proposed FF relay yields a considerable gain over the AF relay. However, the FF relay shows a performance loss when compared to the OFDM-processing method. This loss results from the incapability of the FF relay of subcarrier reordering and the lack of the knowledge of the RD channel state.

\begin{figure}[http]
\centerline{
\SetLabels
\L(0.25*-0.1) (a) \\
\L(0.75*-0.1) (b) \\
\endSetLabels
\leavevmode
\strut\AffixLabels{
\scalefig{0.5}\epsfbox{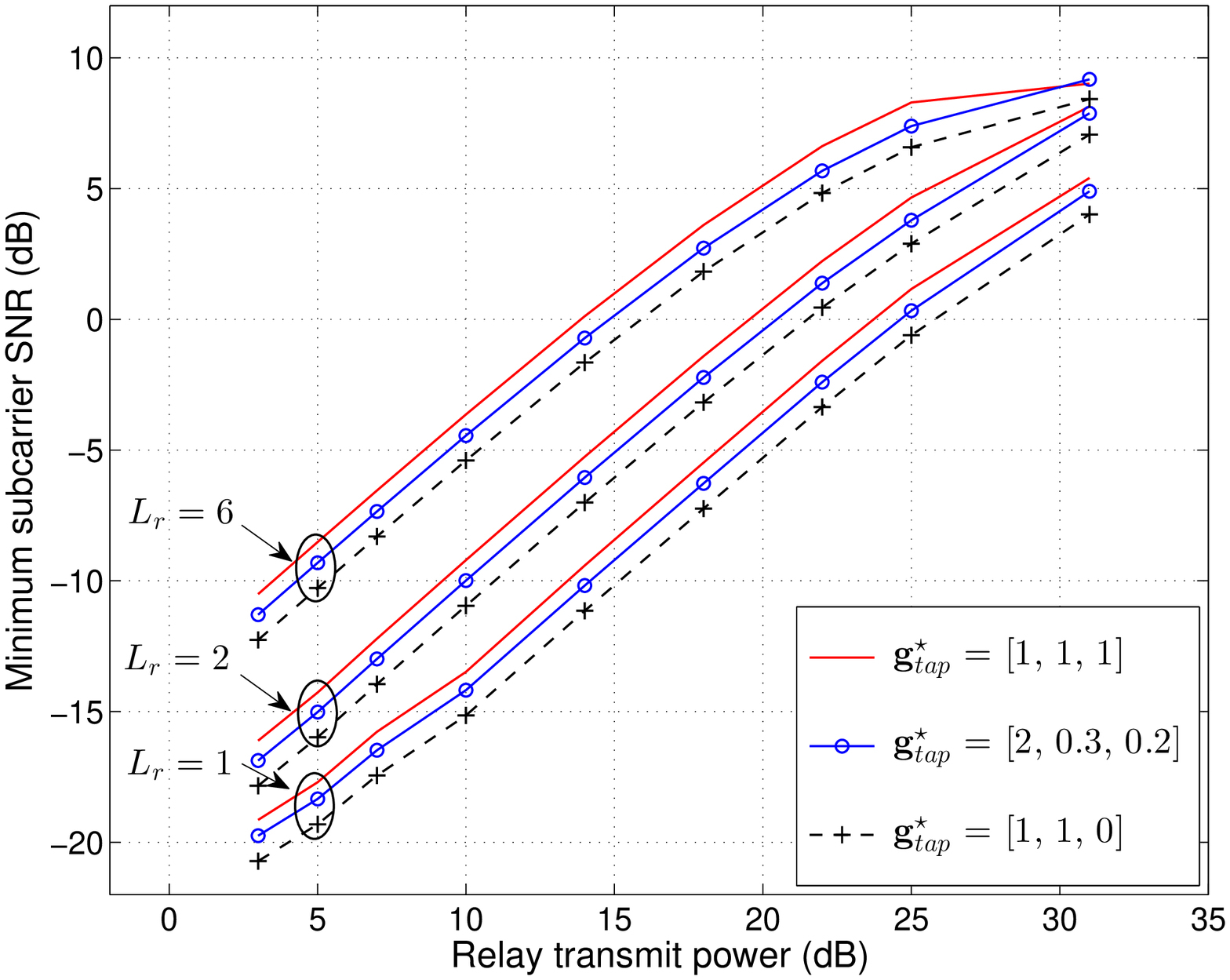}
\scalefig{0.5}\epsfbox{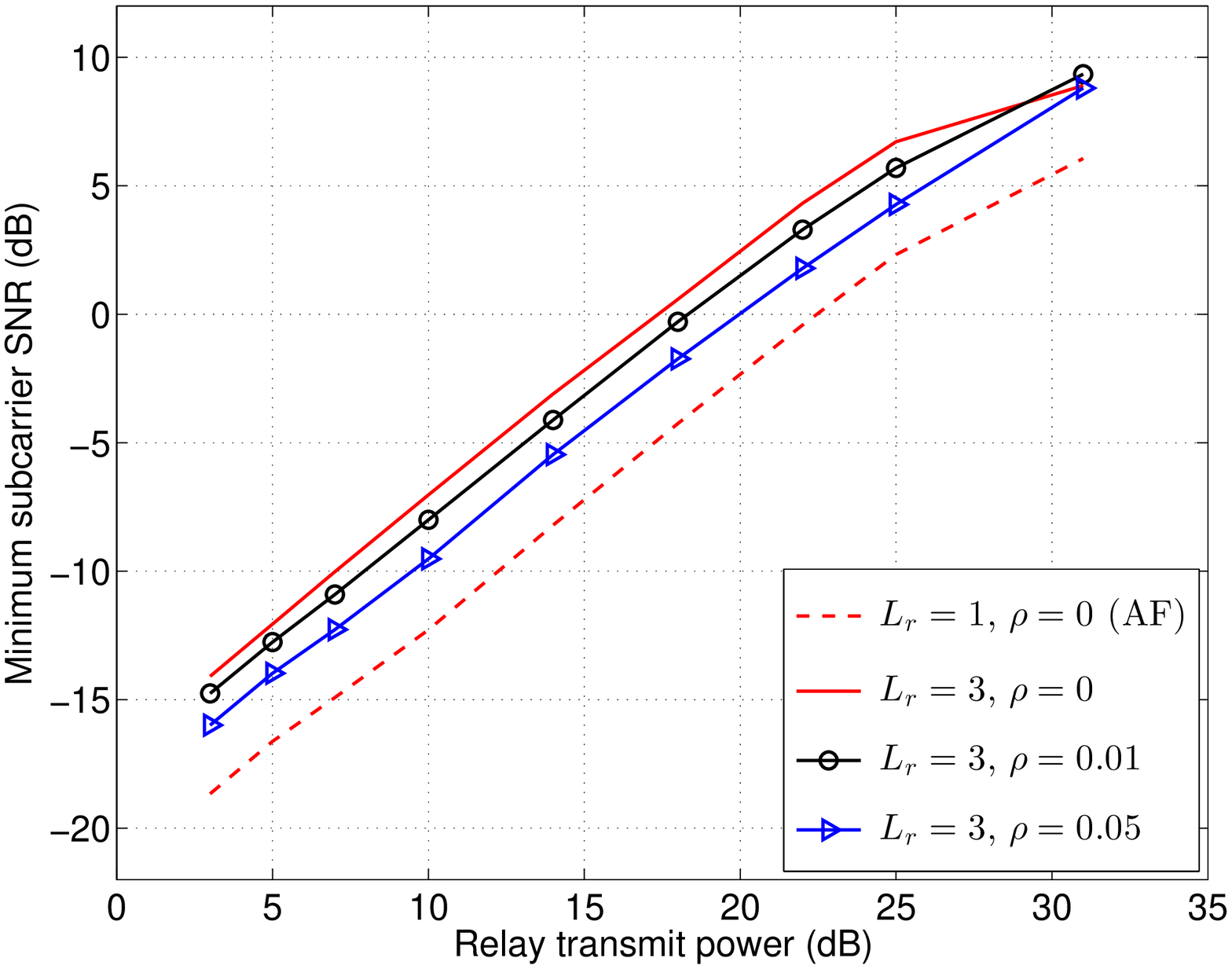} } }
\vspace{1.5em}
\centerline{
\SetLabels
\L(0.25*-0.1) (c) \\
\L(0.75*-0.1) (d) \\
\endSetLabels
\leavevmode
\strut\AffixLabels{
\scalefig{0.5}\epsfbox{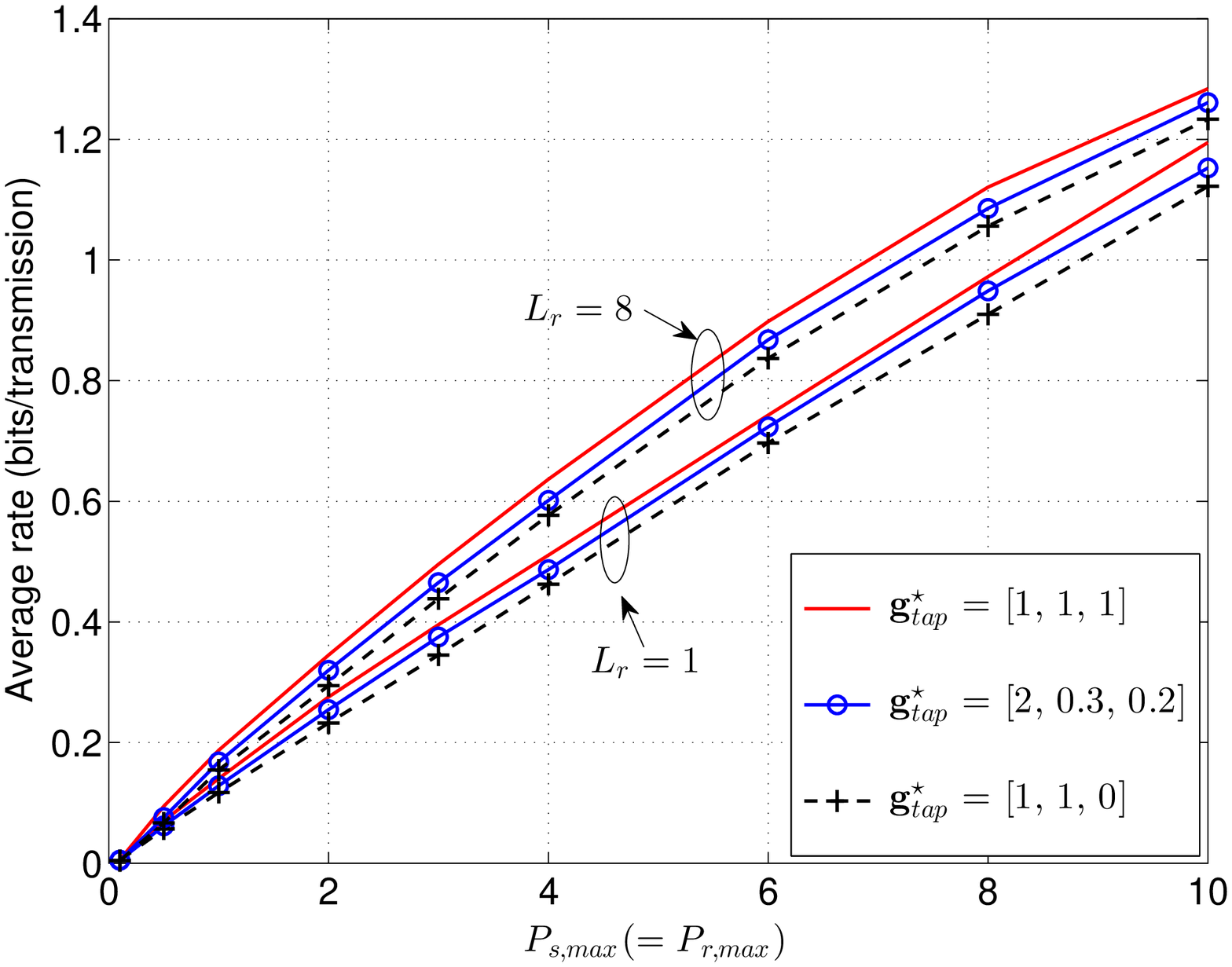}
\scalefig{0.5}\epsfbox{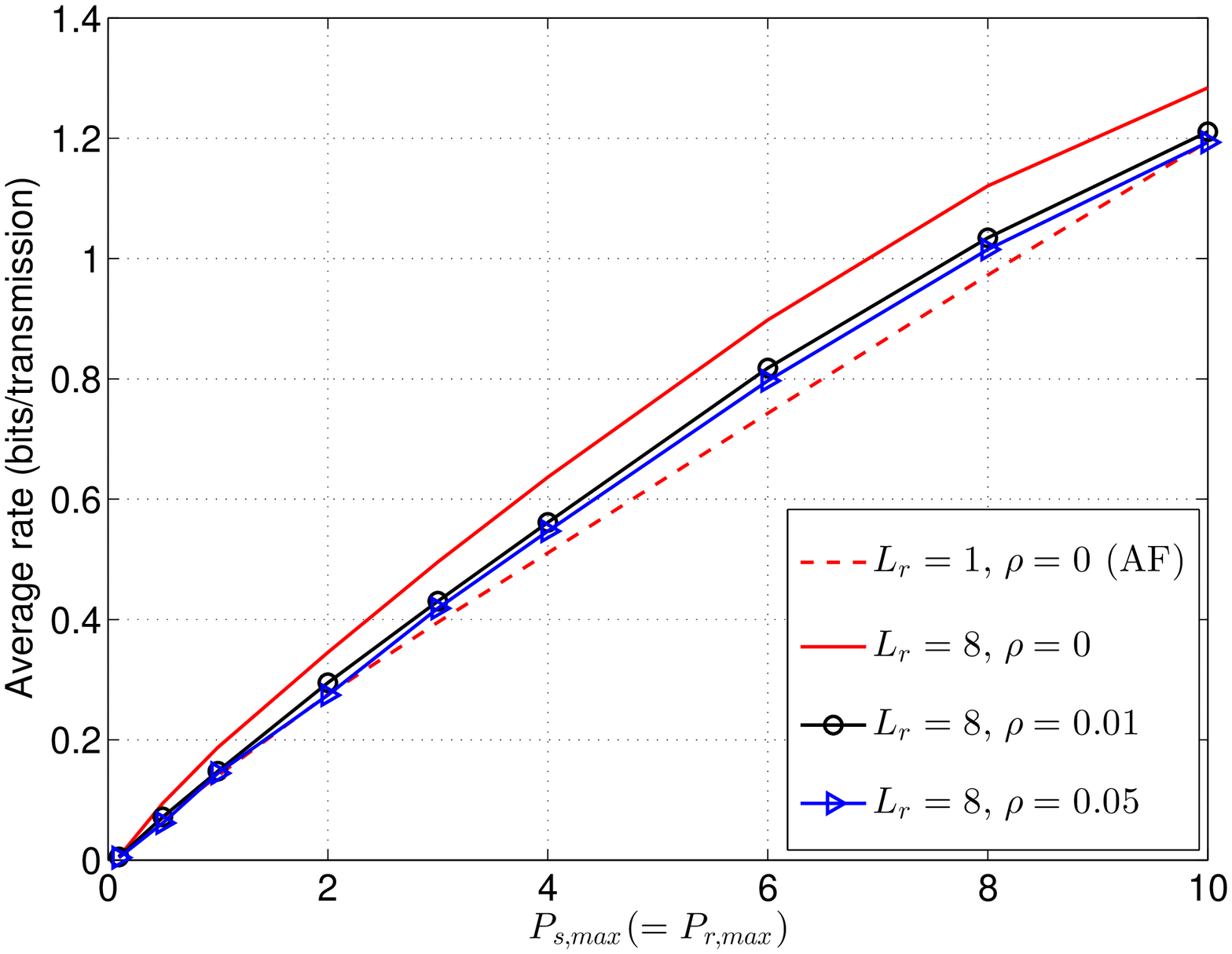} } }
\vspace{1.5em} \caption{$P_{s,max}=P_{r,max}=20$ dB, $L_f=L_g=3$: (a) the worst subcarrier SNR versus $P_{r,max}$ (Algorithm 2-the RD channel statistic mismatch only), (b) the worst subcarrier SNR versus $P_{r,max}$ (Algorithm 2-the SR state mismatch only), (c) the rate versus $P_{s,max}=P_{r,max}$ (The proposed rate maximization algorithm-the RD channel statistic mismatch only), and (d) the rate versus $P_{s,max}=P_{r,max}$ (The proposed rate maximization algorithm-the SR channel state mismatch only)} \label{fig:channelMismatch}
\end{figure}

Finally, we 
examined the robustness of the proposed FF relay design methods, Algorithm 2 and the proposed rate maximization algorithm,
 against channel information mismatch.   {We considered two types of channel information error again with $L_f=3$.  One is the RD channel statistic mismatch and the other is the SR channel state mismatch.} For the RD channel statitic mismatch, still the i.i.d. RD channel model with $L_g=3$ and $\sigma_g^2=1$ for
all taps was used  to run Algorithm 2 and the proposed rate maximization algorithm, but the true RD channel was generated
randomly according to a different channel statistic, i.e., different
$L_g$ and/or different channel power profile. {For the SR channel state mismatch, we modeled  the available information $\hat{\fbf}$ for the SR channel $\fbf$  as
$\hat{\fbf}= \fbf + \Delta\fbf$,
 where  
 $\Delta \fbf=[\Delta f_0,\Delta  f_1,\cdots,\Delta f_{L_f-1}]^T$ is the channel information error vector. Here, the true channel coefficient $f_i$ was generated i.i.d. according to  ${\cal{CN}}(0, 1)$ for $l = 0, 1, \cdots,
L_f-1$, as mentioned already, and the channel information error $\Delta f_i$ was generated i.i.d. according to   $\Delta f_l
\stackrel{i.i.d.}{\sim} {\cal{CN}}(0, \rho)$ for $l = 0, 1, \cdots,
L_f-1$. Thus, $\rho$ is the relative power of the channel information error to the power of the true channel tap.} 
Fig. \ref{fig:channelMismatch} (a) 
 shows the worst subcarrier SNR obtained by Algorithm 2 (averaged 500 over
channel realizations) versus the relay transmit power in the case
of RD channel static information mismatch and correct SR channel state information.  
It is seen  that the proposed FF relay design method
is robust against the channel static mismatch.  {Fig. \ref{fig:channelMismatch} (b) 
shows the impact of the SR channel mismatch on Algorithm 2 with the correct RD static information. 
As expected, the AF case $L_r =1$ is most robust against the SR channel information error since it has only one tap. As the number of the FF filter taps increases, there is  noticeable performance degradation in the case of the SR channel static mismatch.}  {Figures  \ref{fig:channelMismatch} (c) and (d) show the impact of the RD channel static mismatch and the SR channel state mismatch on the proposed rate maximization algorithm, respectively. Similar behavior is seen as in the worst subcarrier SNR maximization.} {In both cases, the algorithms are more robust against the RD static mismatch than against the SR state mismatch. Thus, accurate channel estimation of the SR state is necessary.  Fortunately, in most cellular communication systems, there exist  pilot signals from the basestation which can be used for channel estimation and the SR channel state can be estimated accurately at the relay by using the pilot signals.}

\section{{Discussion}}
\label{sec:discussion}

In this section, we discuss several practical issues to implement the proposed full-duplex FF relay. First, let us consider the full-duplex operation.  The main advantage of AF relays (i.e., simple reapters) is that they can be operated in the full-duplex mode, and this full-duplex operation incurs no rate reduction inherent to half duplexing. In the full-duplex operation, however,  we have the  problem of self-interference; i.e., the transmitted signal from the relay is fed back to the receiver of the relay. However, this self-interference problem already exists with full-duplex AF relays. There exists vast literature on mitigation of self-interference for AF relays \cite{Nasr&Cosmas&Bard&Gledhill:07BC, Slingsby&McGeehan:95, Hamazumietal:03, Anderson&Krishnamoorthy:04}.  It is shown in \cite{Nasr&Cosmas&Bard&Gledhill:07BC} that echo cancellation  combined with the physical separation of transmission and reception antennas at the relay can effectively solve the self-interference problem of full-duplex relays.  In the case that  interference cancellation is employed at AF relays already, the additional processing for full-duplex FF over full-duplex AF is insignificant because of  the already existing up and down conversion and baseband processing for echo cancellation for full-duplex AF relays.

 Next consider the availability of the channel  information assumed in the previous sections. 
 In many research works for relays, it is assumed that all channel information  is available at the transmitter and the relay. For a non-transparent relay, this assumption is valid  since the relay has its own identity and can transmit its own pilot signal to terminal stations and the relay can get feedback from terminal stations. However, for the cheap transparent operation, the relay does not have a physical identity and does not receive any feedback for terminal stations.  Although the transparent relay is invisible to terminal stastions, there still exists a control communication link between the basestation and the transparent relay in real world systems for maintenance purposes; Basic relay operation commands from the basestation should be delivered to the relay and operation condition information should be fed back to the basestation from the relay. 
In addition to this basestation-relay control communication link, there exists a low-rate robust control link from terminal stations to the basetation in all cellular networks.\footnote{The control channel typically operates at a low rate. It compensates low signal power with long bit duration. Thus, although the direct link from the basestation to the terminal station  is seriously faded, we can still assume that the control link operates properly.} Typically, through this link, channel quality indication (CQI) and/or channel state information (CSI) is fed back to the basestation. 
Exploiting these two control links and the (typically existing) pilot signal from the basestation, one can estimate the necessary channel information assumed in the previous sections, as follows. 

\vspace{0.5em}

Step 1. From the pilot  signal $x_p[n]$ transmitted from the basestation, the relay estimates the SR channel state $f[l]$ immediately. For example, one can use preamble signals attached in time domain to OFDM signals. There are several time-domain channel estimation techniques  for OFDM signals not requiring ODFM processing.

 Step 2. The relay filters the incoming signal with the FIR response $r[l]$ and transmits the filtered signal to the destination. 

 Step 3. The destination does not know the existence of the relay in the transparent mode, but what the destination receives for the pilot portion is $y_d[n] = x_p[n] * f[n] * r[n] * g[n]$ under the assumption that the SD channel strength is negligible. As usual, the destination node estimates the channel $h[l]$ based on the pilot signal $x_p[n]$. The estimated channel at the relay is then $h[l]= f[l]* r[l] * g[l]$.

 Step 4. As in most cellular systems, the destination node feedbacks the CSI $h[l]$ to the basestation via the available uplink control channel.

 Step 5. The FF relay also feedbacks the SR CSI $f[l]$ and its filter response $r[l]$ to the basestation via the available control channel between the basestation and the relay.

 Step 6. The basestation now has $f[l]$, $r[l]$ and $h[l]=f[l] * r[l] * g[l]$. It can simply deconvolve $f[l]*r[l]$ from  $h[l]$ to obtain $g[l]$. In this stage, it seems more practical and robust to extract and use  the stastic of $g[l]$ as in this paepr since the CSI is prone to phase errors in the RF circuitries at the relay and communication delays.  For example, the delay spread and the channel gain magnitude information can be extracted as in the previous sections. 

 Step 7. The basestation computes the FF relay filter response $r[l]$ (based on the results in the previous sections) and downloads the information to the relay via the control channel between the basestation and the relay. In this case, the information $r[l]$ does not need to be fed back to the base station from the relay since the base station already has $r[l]$. In this way, computational burden is moved to the basestation and this strategy seems reasonable for the joint optimization considered in Sections \ref{subsec:worstsubSNRmax} and
\ref{subsec:ratemax}.

\vspace{0.5em}  As seen in the above, a practical implementation of the proposed FF scheme is possible and does not require any standard change. 

Finally, we consider the possibility of extension of the FF relay to the broadcasting situation in which the basestation serves several terminal stations in the relay cell simultaneously. In this paper,  we assumed channel state information for the SR
channel and channel statistic information for the RD channel. The assumption of
channel statistic information for the RD channel makes the
proposed FF relay design method useful for the broadcasting
purpose. Suppose that the basestation acquires the channel information from each terminal station  by the method in the above paragraph. Then, the basestation has the RD channel information from all terminal stations in the relay cell. The basestation can select and schedule terminal stations with similar channel statistics and designs the relay filter accordingly. In this way, the proposed FF relay scheme can be applied to the broadcasting scenario since 
the proposed design methods do not require exact CSI or the phase of the channel gain for the RD channel and are robust against the RD channel static mismatch. In this broadcasting scenario, the worst subcarrier SNR maximization in Section \ref{subsec:worstsubSNRmax} improves QoS fairness among users and the rate maximization in 
Section \ref{subsec:ratemax} increases the overall system sum rate.

\section{Conclusion}
\label{sec:conclusion} 

In this paper, we have considered the FF relay design for OFDM
systems for transparent relay operation to compromise the performance and complexity between the simple repeater and the full OFDM-processing relay. {
We have considered three FF relay design criteria of minimizing the relay transmit power subject to per-subcarrier
SNR constraints,  maximizing the worst subcarrier SNR subject to transmit power constraints, and maximizing the data rate subject to transmit power constraints. We have proposed an efficient algorithm for each of the three criteria based on convex relaxation, alternating optimization and the projected gradient method.} The proposed FF relay significantly outperforms the simple repeater with slight increase in complexity and the same operating condition, and thus provides an effective alternative to the simple repeater. 
In this paper, we assumed SISO-OFDM systems. However,
most current OFDM systems employ MIMO communications and thus,
extension to the MIMO case is left as a future work.

\section*{Appendix}

 {\it Proof of Theorem \ref{theo:OptimalityRelaxedSDP}}

By introducing a slack variable $\tau$, we convert Problem
1$^\prime$ to the following equivalent problem:
\begin{eqnarray}
& \underset{\tau, \Rc}{\min} & \tau   \label{eq:AppendProbSlack}\\
& {\mbox{s.t.}}  & \mbox{tr} ( \Phibf_P \Rc ) \le \tau,  \label{eq:AppendLagCon1}\\
& & {\mbox{tr}} ( [\Phibf_S (k) -\gamma_k \Phibf_N(k)] \Rc
 )\ge \sigma^2_d \gamma_k, ~~~~~  k = 0, 1, \ldots,  N-1,
 \label{eq:AppendLagCon2}\\
 & & \tau \ge 0, \label{eq:AppendLagCon3}\\
 & &  \Rc \succeq 0. \label{eq:AppendLagCon4}
\end{eqnarray}
The Lagrange dual function for the above problem is given by
{\small
\begin{equation}
 g(\lambda, \{\mu_k\}, \nu, \Psibf ) = \inf_{\tau, \Rc} \left( (1-\lambda-\nu) \tau
 + \sum_{k=0}^{N-1} \mu_k \gamma_k \sigma_d^2 + \mbox{tr}
 ( \{ ~\underbrace{ \lambda \Phibf_P  - \sum_{k=0}^{N-1} \mu_k [ \Phibf_S (k) - \gamma_k \Phibf_N(k)]}_{=:\Qbf(\lambda, \{\mu_k\})} -\Psibf\} \Rc
 ) \right),
\end{equation}}
where $\lambda \ge 0$,   $\{\mu_k \ge 0\}$, $\nu \ge 0$ and
$\Psibf \succeq 0$ are the dual variables associated with
\eqref{eq:AppendLagCon1}, \eqref{eq:AppendLagCon2},
\eqref{eq:AppendLagCon3}, and \eqref{eq:AppendLagCon4},
respectively. If $1-\lambda-\nu \ne 0$ or
$\Qbf(\lambda,\{\mu_k\})-\Psibf \ne 0$, then the dual function
value is minus infinity or we have trivial solutions $\tau=0$
and/or $\Rc ={\mathbf{0}}$. Thus, for the nontrivial feasibility
of $\tau$ and $\Rc$, we have $1-\lambda =\nu ~(\ge 0)$ and
$\Qbf(\lambda, \{\mu_k\}) = \Psibf ~(\succeq 0)$. Then, the
Lagrange dual function is easily obtained as $g(\lambda,
\{\mu_k\}, \nu, \Psibf )=\sum_{k=0}^{N-1} \mu_k \gamma_k
\sigma_d^2$ and the corresponding dual problem is given by
\begin{eqnarray}
& \underset{\lambda, \{\mu_k\}}{\max} & \sum_{k=0}^{N-1} \mu_k \gamma_k \sigma_d^2  \nonumber\\
& {\mbox{s.t.}}  & 0 \le  \lambda \le 1, ~~
\Qbf(\lambda,\{\mu_k\}) \succeq 0, ~~\mu_k \ge 0, ~~  k = 0, ~1,~
\ldots,~ N-1.
\end{eqnarray}
 Let
$\lambda^\star,~ \{\mu_k^\star\},~ \tau^\star, ~\Qbf^\star$ and $
\Rc^\star$ be the optimal values for the problem. ($\nu^\star$ and
$\Psibf^\star$ are automatically determined based on these
quantities. The dependence of $\Qbf$ on $\lambda$ and $\{\mu_k\}$ is
not shown explicitly for notational simplicity from here on.) From
the complementary slackness conditions for \eqref{eq:AppendLagCon1}
and \eqref{eq:AppendLagCon2}, we have
\begin{equation} \label{eq:SumofComplementarySlackness}
 \lambda^\star \left( \mbox{tr}\left( \Phibf_P \Rc^\star \right) -
 \tau^\star \right) + \sum_{k=0}^{N-1} \mu_k^\star  \left( \gamma_k\sigma_d^2 - \mbox{tr}
  \left( [\Phibf_S (n) - \gamma_k \Phibf_N (k) ] \Rc^* \right)\right)
  =0,
\end{equation}
which is equivalent to
\begin{equation} \label{eq:SumofComplementarySlackness2}
\left(\sum_{k=0}^{N-1} \mu_k^\star  \gamma_k\sigma_d^2 -
\lambda^\star \tau^\star \right) +  \tr \left( \{
\underbrace{\lambda^\star \Phibf_P - \sum_{k=0}^{N-1} \mu_k^\star
[\Phibf_S (k) - \gamma_k \Phibf_N (k) ]}_{=\Qbf^\star}\} \Rc^\star
\right)
  =0.
\end{equation}
Since the problem
(\ref{eq:AppendProbSlack}-\ref{eq:AppendLagCon4}) is a convex
optimization problem, the duality gap is zero, i.e.,
$\sum_{k=0}^{N-1} \mu_k^\star  \gamma_k\sigma_d^2 = \tau^\star$.
Thus, both the first and second terms in the left-hand side (LHS)
of \eqref{eq:SumofComplementarySlackness2} are nonnegative since
$\sum_{k=0}^{N-1} \mu_k^\star  \gamma_k\sigma_d^2 - \lambda^\star
\tau^\star= \tau^\star (1-\lambda^\star) \ge 0$ and
$\tr(\Qbf^\star \Rc^\star) \ge 0$. (The trace of the product of
two positive semi-definite matrices is nonnegative
\cite{Helmberg:02SDP}.) Therefore,  $\lambda^\star = 1$ and
$\mbox{tr} ( \Qbf^\star \Rc^\star  )=0$. It is obvious that
$\Qbf^\star\nsucc 0$ for a nontrivial $\Rc^\star$ from $\mbox{tr}
( \Qbf^\star \Rc^\star )=0$, i.e., the $L_r \times L_r$ matrix
$\Qbf^\star$ does not have full rank. This is because $\mbox{tr} (
\Qbf^\star \Rc^\star ) = \sum_i \sigma_i\mbox{tr}(\Qbf^\star
\ubf_i\ubf_i^H)= \sum_i \sigma_i(\ubf_i^H\Qbf^\star \ubf_i)$,
where $\Rc =\sum_i \sigma_i\ubf_i\ubf_i^H$  is the
eigen-decomposition of $\Rc^\star$. (If $\Qbf^\star\succ 0$, then
$\mbox{tr} ( \Qbf^\star \Rc^\star  ) > 0$.)  Note from
\eqref{eq:SumofComplementarySlackness2} that
\begin{equation}
\Qbf^\star =  \Phibf_P - \sum_{k=0}^{N-1} \mu_k^\star \left(
\Phibf_S (k) - \gamma_k \Phibf_N(k)\right),
\end{equation}
where $\Phibf_P$ is a positive definite matrix defined in
(\ref{eq:RelayPower}), $\Phibf_N(k) $ is a positive semi-definite
matrix defined in (\ref{eq:ydnk_noise_power}), and $\Phibf_S (k)$
defined in (\ref{eq:ydsk_sig_final}) is a rank-one matrix by Lemma
\ref{lemma:Append}.  Now, under the assumption that all the SNR
constraints except one are satisfied with strict inequality, we
have $\mu_i \neq 0$ for some $i$ and $\mu_j = 0 ,~ \forall j \neq
i $ from the complementary slackness conditions. In this case,
$\Qbf^\star$ is given by
\begin{equation} \label{eq:Qmatrix}
 \Qbf^\star = \underbrace{ \Phibf_P +  \mu_i^\star \gamma_{i} \Phibf_N(i)}_{\mbox{rank $L_r$}} -  \underbrace{\mu_i^\star \bs{\Phi}_S (i)}_{\mbox{rank $1$}}.
\end{equation}
Due to the structure of $\Qbf^\star$ in \eqref{eq:Qmatrix}, the
rank of  $\Qbf^\star$ is larger than or equal to $L_r -1$. Since
$\Qbf^\star\nsucc 0$,
 ${\mbox{rank ($\Qbf^\star$)} = L_r-1}$.
Since $0=\mbox{tr} ( \Qbf^\star \Rc^\star  )=  \sum_{i=1}^{L_r-1}
\eta_i(\vbf_i^H\Rc^\star \vbf_i)$ (where $\Qbf^\star =
\sum_{i=1}^{L_r-1}\eta_i \vbf_i\vbf_i^H$ is the
eigen-decomposition of $\Qbf^\star$), we conclude that $\Rc^\star$
has nullity $L_r-1$ and thus has rank one under the assumption of
Theorem \ref{theo:OptimalityRelaxedSDP}. \hfill{$\blacksquare$}

\begin{lemma} \label{lemma:Append}
If  $N - L_f +1 > L_f + L_g -1$, $\Phibf_S(k)$ has rank one
regardless of the value of $k$.
\end{lemma}

\vspace{1em}

 {\it Proof of Lemma \ref{lemma:Append}:} Recall  that
(see \eqref{eq:SignalPower} and \eqref{eq:RelayPowerSDP})
\begin{equation}
(NP_{s,k}\sigma_g^{2})^{-1}\Phibf_S (k) = \Ebf_1 \tilde{\Kbf}_k
\Ebf_1^H, ~~~\tilde{\Kbf}_k =  \tilde{\Ibf}_{L_g} \otimes \Kbf_k,
~~~\Kbf_k= \Fbf^* \Tbf^* \wbf_k \wbf_k^H \Tbf^T \Fbf^T.
\end{equation}
Let $\Ebf_1$ in \eqref{eq:E1} be partitioned as $ \Ebf_1 =   \left
[ \Ebf_1^{(1)} , ~ \Ebf_1^{(2)}, ~\ldots , ~\Ebf_1^{(N+L_g-1)}
\right]$. Note that $\Fbf$ and $\Tbf$ are Toeplitz matrices and
that $\wbf_k \wbf_k^H$ is also a Toeplitz matrix regardless of $k$
due to the property of DFT matrices. It is not difficult to show
that $\Kbf_k (1:N-L_f+1,1:N-L_f+1)$ is a Toeplitz matrix, where
$\Abf(a:b, c:d)$ denotes a submatrix of $\Abf$ composed of the
rows from $a$ to $b$ and columns from $c$ to $d$. Now, $\Phibf_S
(k)$ can be rewritten as
\begin{equation} \label{eq:PhiS}
(NP_{s,k}\sigma_g^{2})^{-1}\Phibf_S (k) = \Ebf_1^{(1)} \Kbf_k
\Ebf_1^{(1)H} + \Ebf_1^{(2)} \Kbf_k \Ebf_1^{(2)H} +   \cdots +
\Ebf_1^{(L_g)} \Kbf_k \Ebf_1^{(L_g)H}.
\end{equation}
Here,  the operation $ \Ebf_1^{(i)} \Kbf_k \Ebf_1^{(i)H}$ extracts
a $L_r \times L_r $ submatrix $\Kbf_k(i : L_r + i -1 , ~i : L_r +
i -1)$ from  $\Kbf_k$. If $N-L_r+1 > L_r+L_g-1$, this operation
extracts the same submatrix from $\Kbf_k$ regardless of $i$ since
$\Kbf_k (1:N-L_r+1,1:N-L_r+1)$ is a Toeplitz matrix. Thus, we have
\begin{equation}
(NP_{s,k}\sigma_g^{2})^{-1}\Phibf_S (k) =  L_g \Ebf_1^{(1)} \Kbf_k
\Ebf_1^{(1)H}  = L_g (\Ebf_1^{(1)}\Fbf^* \Tbf^* \wbf_k)
(\Ebf_1^{(1)}\Fbf^* \Tbf^* \wbf_k)^H,
\end{equation}
and $\Phibf_S (k)$ has rank one if the condition $N-L_f+1
> L_t+L_g-1$. \hfill{$\blacksquare$}



\end{document}